\documentclass[preprint]{aastex61}
\usepackage{amsmath,amstext}
\usepackage[T1]{fontenc}
\usepackage{apjfonts} 
\usepackage[figure,figure*]{hypcap}
\usepackage{makecell}
\usepackage{mathrsfs}
\usepackage{mathtools}

\usepackage{color}

\newcommand{\etal}{\textit{et al.}}

\newcommand{\Br}{\rm{B}}

\received{XXX}
\revised{XXX}
\submitjournal{ApJS}

\shorttitle{Morphology Classification of Radio AGNs with CNN}
\shortauthors{Ma \etal}

\begin{document}

\title{A Machine Learning Based Morphological Classification of 14,245 Radio AGNs Selected From The Best--Heckman Sample}

\correspondingauthor{Haiguang Xu}
\email{hgxu@sjtu.edu.cn}

\author{Zhixian Ma}
\affiliation{Department of Electronic Engineering, Shanghai Jiao Tong University,
  800 Dongchuan Road, Minhang, Shanghai 200240, China}

\author{Haiguang Xu}
\affiliation{School of Physics and Astronomy/Tsung-Dao Lee Institute, 
 Shanghai Jiao Tong University, 800 Dongchuan Road, Minhang, Shanghai 200240, China}
\affiliation{IFSA Collaborative Innovation Center, Shanghai Jiao Tong University, Minhang, Shanghai 200240, China}
\affiliation{Department of Astronomy, Shanghai Jiao Tong University,
  800 Dongchuan Road, Minhang, Shanghai 200240, China}
  
\author{Jie Zhu}
\affiliation{Department of Electronic Engineering, Shanghai Jiao Tong University,
  800 Dongchuan Road, Minhang, Shanghai 200240, China}
  
\author{Dan Hu}
\affiliation{Department of Astronomy, Shanghai Jiao Tong University,
  800 Dongchuan Road, Minhang, Shanghai 200240, China}
  
\author{Weitian Li}
\affiliation{Department of Astronomy, Shanghai Jiao Tong University,
  800 Dongchuan Road, Minhang, Shanghai 200240, China}

\author{Chenxi Shan}
\affiliation{Department of Astronomy, Shanghai Jiao Tong University,
  800 Dongchuan Road, Minhang, Shanghai 200240, China}

\author{Zhenghao Zhu}
\affiliation{Department of Astronomy, Shanghai Jiao Tong University,
  800 Dongchuan Road, Minhang, Shanghai 200240, China}

\author{Liyi Gu}
\affiliation{SRON Netherlands Institute for Space Research, Sorbonnelaan 2, 3584 CA Utrecht, the
Netherlands}
  
\author{Jinjin Li}
\affiliation{Key Laboratory of Thin Film and Microfabrication Technology (Ministry of Education), Department of Micro/Nano Electronics, Shanghai Jiao Tong University, Shanghai 200240, China}

\author{Chengze Liu}
\affiliation{Department of Astronomy, Shanghai Jiao Tong University,
  800 Dongchuan Road, Minhang, Shanghai 200240, China}

\author{Xiangping Wu}
\affiliation{National Astronomical Observatories, Chinese Academy of Sciences, 20A Datun Road, Beijing 100012, China}

\begin{abstract}
We present a morphological classification of 14,245 radio active galactic nuclei (AGNs) into six types, i.e., typical Fanaroff--Riley Class I / II (FRI/II), FRI/II-like bent-tailed, X-shaped radio galaxy, and ringlike radio galaxy, by designing a convolutional neural network (CNN) based autoencoder, namely MCRGNet, and applying it to a labeled radio galaxy (LRG) sample containing 1442 AGNs and an unlabeled radio galaxy (unLRG) sample containing 14,245 unlabeled AGNs selected from the Best--Heckman sample. We train MCRGNet and implement the classification task by a three-step strategy, i.e., pre-training, fine-tuning, and classification, which combines both unsupervised and supervised learnings. A four-layer dichotomous tree is designed to classify the radio AGNs, which leads to a significantly better performance than the direct six-type classification. On the LRG sample, our MCRGNet achieves a total precision of $\sim 93\%$ and an averaged sensitivity of $\sim 87\%$, which are better than those obtained in previous works. On the unLRG sample, whose labels have been human-inspected, the neural network achieves a total precision of $\sim 80\%$. Also, using the Sloan Digital Sky Survey (SDSS) Data Release 7 (DR7) to calculate the $r$-band absolute magnitude ($M_\mathrm{opt}$) and using the flux densities to calculate the radio luminosity ($L_\mathrm{radio}$), we find that the distributions of the unLRG sources on the $L_\mathrm{radio}$--$M_\mathrm{opt}$ plane do not show an apparent redshift evolution and could confirm with a sufficiently large sample that there could not exist an abrupt separation between FRIs and FRIIs as reported in some previous works. 
\end{abstract}

\keywords{ 
	catalogs ---
	galaxies: statistics ---
	methods: data analysis --- 
	techniques: miscellaneous ---
	radio continuum: galaxies
}

\section{INTRODUCTION}
The study of radio galaxies and their close relatives, radio-loud quasars and blazars, began in the mid-1950s when the Third Cambridge Catalog (3C; \citealt{Mackay1971}) was created. These sources,  whose radio emission is nonthermal, polarized, and prodigiously powerful with luminosities up to $10^{38-39}$ W in the 10 MHz $-$ 100 GHz band, form an important subclass of the active galactic nucleus (AGN) population. A typical bright radio galaxy appears as an exclusively large elliptical galaxy that exhibits a distinctive radio appearance characterized by twin jets (in some sources, however, only one is visible) and twin lobes or plumes. In many cases these twin structures, which straddle a compact core, may extend out to several hundred kiloparsecs on each side of the galaxy. 

By comparing the distance between the brightest regions on either side of  the galaxy with the entire extent of the source, \citet{Fanaroff1974} classified 57 3C radio-loud galaxies into two broad categories, i.e., the low-luminosity Class I sources (FRIs), which show a decreasing radio emission as the distance from the core increases, and the high-luminosity Class II sources (FRIIs), which show an inverse tendency and often have bright hotspots at the ends of the lobes. In addition to the typical Fanaroff--Riley sources, some morphological subclasses have been defined in succeeding investigations, including (1) narrow-angle tail sources (NATs; \citealt{Rudnick1976}), which usually appear as FRIs bent by the ram pressure of the surrounding intracluster medium in galaxy clusters~\citep{Proctor2011}; (2) wide angle tail sources (WATs; \citealt{Rudnick1976}), which simultaneously possess FRII-like jet-hotspot structures and, connected to the ends of the jets, FRI-like large-scale plumes rather than lobes; (3) X-shaped radio galaxies (XRGs; \citealt{Leahy1992,Cheung2007}), which exhibit a pair of faint lobes (the ``wings") oriented at an angle to another pair of bright active lobes; (4) ringlike radio galaxies (RRGs; \citealt{Proctor2011}); and (5) others, such as hybrid-morphology radio sources (\citealt{Gopal-Krishna2000}) that show FRI and FRII lobe morphologies on opposite sides. 

As of today, however, understanding of this taxonomy is inadequate and so is the relevant sample study. For example, \citet{Capetti2017b} found that although the average radio luminosity of FRIIs is higher than that of FRIs, most of the FRIIs in their sample possess a radio power lower than the FRI--FRII transition found in 3C, which is $2 \times 10^{25}$ W Hz$^{-1}$ at 178 MHz ($H_{\rm 0}$ = 50 km s$^{-1}$ Mpc$^{-1}$; \citeauthor{Fanaroff1974}). ~\citet{Leahy1992}, \citet{Cheung2007}, and  \citet{Dennett2002} also reported that the radio luminosities of XRGs, the majority of which possess an FRII morphology, straddle the Fanaroff--Riley transition. These make the problem of the FRI-FRII dichotomy, one of the unresolved mysteries in radio astronomy, more puzzling. 

At present the purpose of the morphological study of radio galaxies is twofold. First, the diversity of the appearances of radio galaxies is a reflection of the physics behind, with which we can investigate the central black holes and their accretion modes, interactions between radio galaxies and surrounding gas, AGN feedback mechanisms, and so on, when complemented by requisite spectral information (see reviews of \citealt{McNamara2007} and~\citealt{Fabian2012}). Second, radio galaxies make a significant contribution to the extragalactic emission in the low-frequency radio sky, causing a severe impediment to recovering the redshifted 21 cm signals of neutral hydrogen from the Epoch of Reionization (EoR), a period yet to be explored during which the universe was reionized by the emissions of the first-generation objects. According to the estimations of, e.g., \citet{Mellema2013} and \citet{Chapman2016}, the total emission of radio galaxies in $50-200$ MHz is more luminous than that of the redshifted EoR signals by 2--3 orders of magnitude. In direct imaging studies and studies of two-dimensional power spectra in terms of EoR window analysis (e.g., \citealt{Chapman2016}), the situation is further complicated by the diverse morphologies of radio galaxies, because even subpercent-level residuals in removing the radio galaxy component from the radio maps can ruin any attempt to reveal the EoR signals (e.g., \citealt{Beardsley2016,Chapman2016,Procopio2017}). Thus, when detecting the EoR signals with facilities such as the Square Kilometre Array (SKA; \citealt{Mellema2013,Koopmans2015}) and the Hydrogen Epoch of Reionization Array (HERA; \citealt{DeBoer2017}), the radio galaxy foreground component must be correctly identified and thoroughly removed. 

A tremendously large number of radio galaxies will be detected with the high-quality radio data obtained by SKA, HERA, and their scale-down analogs. For example, from the simulation of the SKA Design Study (\citealt{Wilman2008, Wilman2010}), it is estimated that about $2.5 \times 10^{4}$ radio sources per square degree will be detected at 1.4 GHz with a limit of 1 $\mu$Jy in the SKA-MID deep survey at 350 MHz --14 GHz. This estimation agrees with the results of \citet{Padovani2016} very well, implying a significant increase of the radio galaxy number density compared to that in the Faint Images of the Radio Sky at Twenty cm (FIRST) Data Release 7 (DR7; $\sim 100$ sources per square degree at 1.4 GHz) by a factor of $\sim 10^{3}$.

Given such a large number of sources with various morphologies, it is obvious that traditional human classification, which is primarily based on visual examination, will be impractical. Machine-learning techniques must be summoned to implement source detection and classification tasks in a correct and efficient way. In fact, machine learning has been applied to help analyze stellar spectra (e.g., \citealt{Fabbro2017}), estimate photometric redshifts (e.g., \citealt{Mountrichas2017}), search for signals from pulsars (e.g., \citealt{Bethapudi2017}) and gamma-ray bursts (e.g., \citealt{Ukwatta2016}), classify galaxies according to their morphologies (e.g., \citealt{Proctor2011,Dieleman2015,Barchi2017,Lukic2018}), predict solar flares (e.g., \citealt{Benvenuto2017}), model galaxy formation and evolution (e.g., \citealt{Kamdar2016}), and so on. As reported by \citet{Aniyan2017}, who successfully classified 178 FRI, 284 FRII, and 254 bent-tailed (BT) galaxies, the morphological classification of radio galaxies is also a task potentially well-suited for deep machine learning when the convolutional neural networks (CNNs) are applied (see also \citealt{Lukic2018}).  

CNNs are a class of biologically inspired variants of feedforwarded multilayer perceptrons that have shown excellent performance in imaging analysis due to their distinguishing features, i.e., 3D volumes of neurons, a local receptive field, and shared weights~\citep{Goodfellow2016}. By training a neural network with augmented sample images and applying a fusion classifier based on a majority-voting classifier, \citet{Aniyan2017} found that the neural network could classify FRI and BT images with very high precision ($91\%$ and $95\%$), although for FRII images the precision was relatively lower ($75\%$). They argued that the low precision for FRII might have been caused by the fact that many BT sources exhibited an FRII appearance and were classified as FRIIs.

The primary motivation of this paper is to develop an automatic open-source tool that can classify a large sample of radio galaxies based on morphological representation learning \citep{LeCun2015} with high computational efficiency, small resource consumption, and powerful predictive capabilities, so as to enable relevant studies on portable computers. In order to do this we have designed a convolutional autoencoder (CAE) based on a deep CNN, unsupervisedly pre-trained the CAE with 14,245 unlabeled images of radio AGNs in the Best--Heckman sample (\citealt{Best2012}, BH12 hereafter), and supervisedly fine-tuned the CAE with images of 1442 radio AGNs that have already been labeled in the literature~\citep{Gendre2008, Gendre2010,Capetti2017a,Capetti2017b,Baldi2017,Cheung2007,Proctor2011}. Given that currently no large labeled radio galaxy (LRG) sample is available for neural network training, this approach (i.e., unsupervisedly pre-training the network with a large unlabeled sample and supervisedly fine-tuning it with a relatively small labeled sample) can best extract information hidden in the images and make the classification robust to morphological variability (e.g., \citealt{Erhan2010,Hinton2012,Christodoulidis2016}; see also \citealt{LeCun2015} for a review). Also, the application of a large unlabeled sample helps construct a correct network by avoiding the overfitting problem (e.g., \citealt{Goodfellow2016}). 

To support astronomical studies, besides the types of FR0, typical FRI, typical FRII, and BT (this category includes both NATs and WATs, which are labeled ``FRI-like" or ``FRII-like" in this work), we extend the classification to the types of XRG and RRG. As a result of this work a catalog of 14,245 radio AGNs, the labels of which have been cross-checked by human inspection, is presented.

This paper is organized as follows. In Section~\ref{sec.smp} we show the classification frame, describe our sample selection criteria, and construct the labeled and unlabeled samples. In Section~\ref{sec.cae} we describe the design of the CAE, which is based on a single small-scale CNN. In Section~\ref{sec.result} we pre-train and fine-tune the neural network, classify the radio galaxies, and evaluate the performance of the network by visualizing the extracted features of the labeled samples with $t$-distributed stochastic neighbor embedding ($t$-SNE) and by comparing the results with those obtained with human inspection on the unlabeled samples. In Sections~\ref{sec.dis} and \ref{sec.summary} we discuss and summarize our results. Throughout the paper we quote errors at the 90\% confidence level unless otherwise stated, and adopt cosmological parameters  $H_{\rm 0}$ = 71 km s$^{-1}$ Mpc$^{-1}$ and  $\Omega_m$ = 1 $-$ $ \Omega_{\Lambda}$ = 0.27. 

\section{Classification Tree and Sample Selection\label{sec.smp}}
By consulting the morphological taxonomies of \citet{Proctor2011}, \citet{Padovani2016}, and \citet{Padovani2017}, we design a four-layer dichotomous tree structure classifier (Fig.~\ref{fig.tree}) and use it to classify radio AGNs. Such a binary classification is expected to achieve better performance than direct multiclass classification under the same network structure and parameter settings (e.g.,~\citealt{Bishop2006,Fei2006,Tsoumakas2007,Ma2017gbtsvm}), because the greater the number of model outputs (i.e., classes), the higher the complexity and stability the model should possess (see, e.g.,~\citealt{Tsoumakas2007,Rifkin2004}) as shown in many practical cases, especially on machine-learning models like the support vector machine (SVM; \citealt{Chang2011libsvm,Tsoumakas2007}). Given that a radio AGN showing a compact appearance can be either a genuine FR0 source or an extended source located at a relatively higher redshift (e.g., $z > 0.05$), we divide the compact category into FR0 (i.e., genuine FR0s located at $z \leq 0.05$) and unresolvable (i.e., sources with a compact appearance and located at $z > 0.05$, including genuine FR0s and extended sources whose linear extension is $\leq$ 6.8 kpc or 5$^{\prime\prime}$) subtypes. Considering that BTs, XRGs, and RRGs more or less exhibit FR morphological features~\citep{Proctor2011}, in the extended source category we define the atypical subtype to include BTs (further classified into FRI-like and FRII-like BTs) and irregular sources (i.e., XRGs and RRGs). 

All the images of the radio AGNs included in both the unlabeled sample (unlabeled radio galaxy sample or unLRG sample) and the labeled sample (LRG sample) are drawn from the Very Large Array (VLA) FIRST survey archive at \url{http://sundog.stsci.edu/} in FITS format with a size of $4'.5 \times 4'.5$ ($150 \times 150$ pixels). 

We construct the LRG sample by combining the CoNFIG catalog (\citealt{Gendre2008, Gendre2010}), the FR0CAT catalog (\citealt{Baldi2017}), the FRICAT catalog (\citealt{Capetti2017a}), the FRIICAT catalog (\citealt{Capetti2017b}), and the catalogs of \citet{Cheung2007} and \citet{Proctor2011} (Table~\ref{tab.tab1} and Table~\ref{tab.tab2}). To be specific, we first include all 392 compacts, 284 FRIs, 587 FRIIs, and 430 BTs labeled in the CoNFIG, FR0CAT, FRICAT, and FRIICAT catalogs in the LRG sample. Note, however, that one source in the FRICAT catalog shows compact morphology, three sources in the FRIICAT catalog are X-shaped-like, five FRIs and eight FRIIs in the FRICAT and FRIICAT are bent, and the compact category in the CoNFIG catalog actually contains sources located at relatively higher redshifts, for which it is impossible to determine whether or not they are intrinsically compact with the spatial resolution of FIRST. To handle these special cases we either carry out visual inspection of their images or inspect their redshifts, and classify the sources into { the} FRI-like, FRII-like, XRG, and unresolvable categories accordingly. Next we merge 410 BTs and 65 RRGs in the \citet{Proctor2011} catalog and 100 XRGs in the \citet{Cheung2007} catalog into the LRG sample, thus obtaining a preliminary sample. At the final step, we filter the preliminary sample by excluding sources with particularly low peak-to-average power ratios (PAPRs) smaller than 10.0 dB, sources with ambiguous morphologies, sources that are too extended, and 37 radio galaxies labeled as ``uncertain" in the CoNFIG catalog. The resulting LRG sample consists of a total of 1442 radio AGNs, which span a redshift range of 0.06 < $z$ < 3.90 and a 1.4 GHz flux density range of 6.0 mJy < $S_{\rm 1.4~GHz}$ < $1.5\times10^{4}$ mJy (see Table~\ref{tab.LRG}). 

We construct the unLRG sample, which does not overlap the LRG sample, by selecting sources directly from the large sample published by \citet{Best2012}. In the BH12 sample the authors cataloged 18,286 radio galaxies by cross-matching the optical spectroscopic catalogs produced by the group at the Max Planck Institute for Astrophysics and Johns Hopkins University (\citealt{Brinchmann2004, Tremonti2004}), who carried out their work based on the DR7 (\citealt{Abazajian2009}) of the Sloan Digital Sky Survey (SDSS) spectroscopic sample, with the National Radio Astronomy Observatory (NRAO) VLA Sky Survey (NVSS; \citealt{Condon1998}) and the FIRST survey \citep{Becker1995}, after applying a flux density level of 5 mJy in the NVSS data (see also \citealt{Best2005} and \citealt{Donoso2009} for more technical details). We limit our selection to the BH12 sample sources by applying the following selection criteria: (1) the source is labeled an AGN according to its optical and radio properties, (2) the source is not overextended compared with the size of the image to be analyzed ($140\times140$ { pixels}); (3) the 1.4 GHz flux density of the source is $\geq 5~\mathrm{mJy}$, and (4) the source is not included in the LRG sample. A  maximum offset of $0{''}.96$ \citep{Proctor2011} is guaranteed when cross-matching the sky coordinates of the candidate sources with those of the LRG sources. As a result 14,245 sources, the redshifts and 1.4 GHz flux densities of which range across $0.01 < z < 0.70$ and $5.0~{\rm mJy} < S_{\rm 1.4~GHz} < 2.3 \times 10^{4}~{\rm mJy}$, respectively, are selected to construct the unLRG sample (see Table~\ref{tab.unLRG}). 

\section{Autoencoder Based on a Deep CNN\label{sec.cae}}
 \subsection{Network Architecture\label{sec.NetArch}}
We have designed an autoencoder, namely MCRGNet (Morphological Classification of Radio Galaxy Network), which consists of an encoder block and a decoder block by convention \citep{Goodfellow2016,Yu2014}, based on a deep CNN to learn the morphological features (i.e., the intrinsic representations of radio AGN images) and to precisely reconstruct the original input images by applying the learned image features. 

Many of the state-of-the-art deep neural networks, e.g., AlexNet~\citep{Krizhevsky2012}, VGG~\citep{Simonyan2014}, and GoogLeNet~\citep{Szegedy2015}, have achieved good performance in image classification and object detection competitions (e.g., ILSVRC; \citealt{ILSVRC15}). They have also been applied to different but related tasks through transfer learning (e.g., \citealt{Pan2010};~\citealt{Esteva2017}). Note that although nearly all of the above networks have been trained on image sets drawn from ImageNet~\citep{Deng2009}, which is composed of 14 million images of 1000 generic object classes spanning a wide range of subjects that cover many aspects of either human activity or life and natural phenomena on Earth~\citep{Esteva2017}, no generic object class is directly related to celestial objects \citep{Lukic2018}. Therefore, negative transfer problems~\citep{Pan2010,Mahmud2008}, e.g., dissimilar risks (see \citealt{Rosenstein2005} for details), may be caused when these networks are directly applied to the task of radio AGN classification for transfer learning. In addition to this, all of these networks are composed of tens of layers, which contain millions of parameters; thus they must be run on workstations with multiple graphics processing units (GPUs) to be trained. In order to consume fewer device resources and enable relevant studies even on portable computers, we have decided to design a small-scale neural network, instead of using the above large networks~(e.g., \citealt{Lukic2018,Diamantis2018}).

\begin{figure}[]
\centering
\gridline{\fig{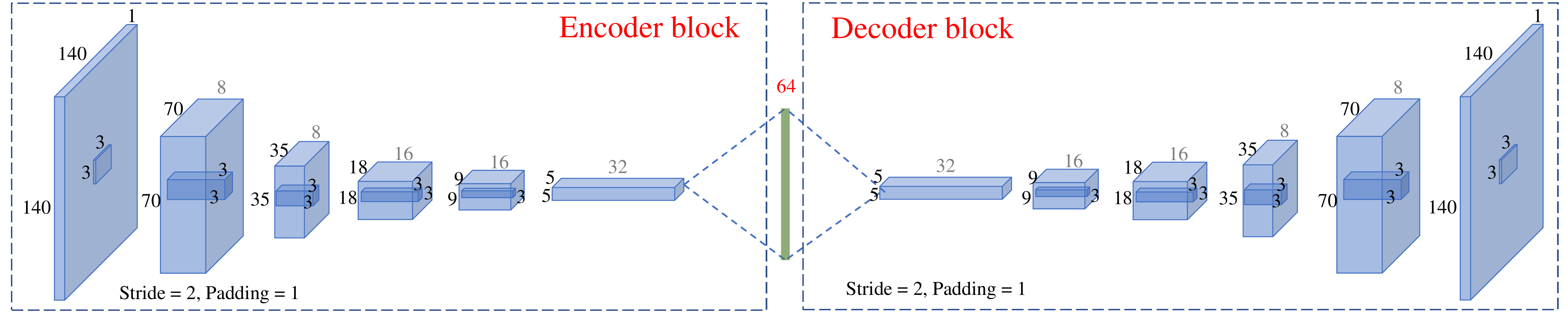}{0.96\textwidth}{(a)\label{fig.cae}}}
\gridline{\fig{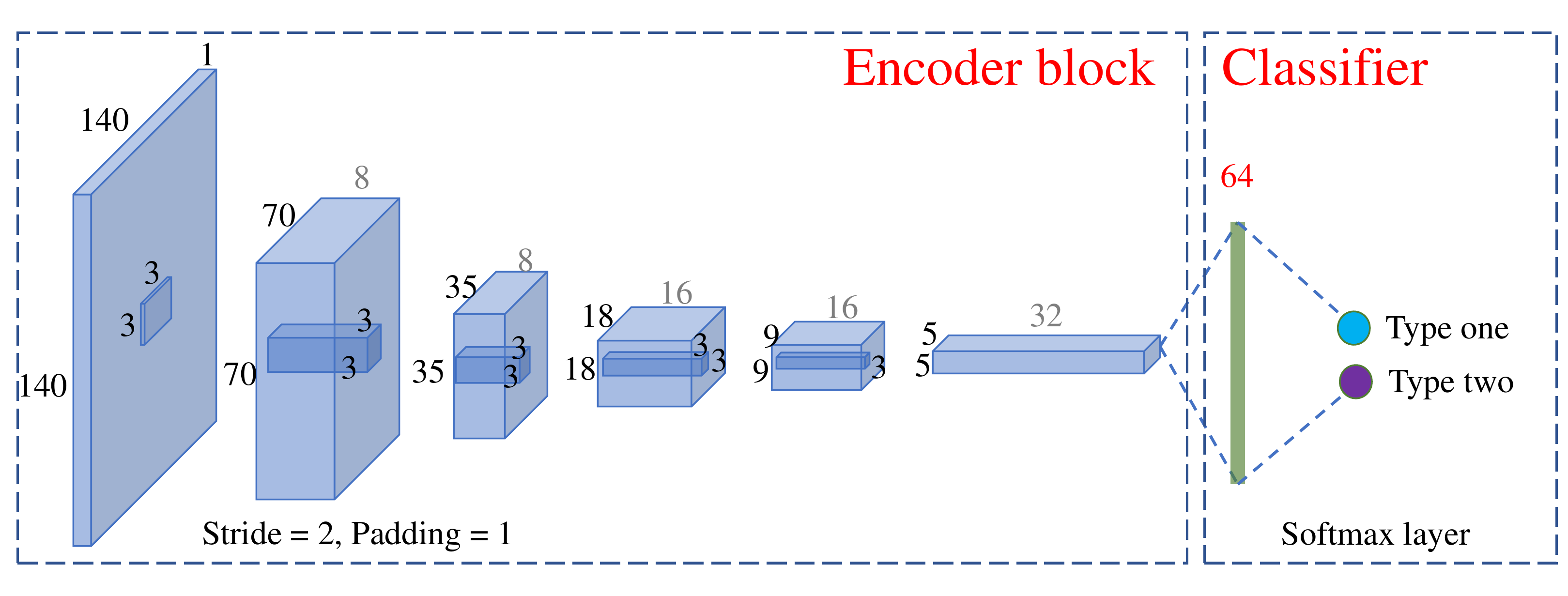}{0.72\textwidth}{(b)\label{fig.cnn}}}
\caption{Schematic diagrams of  
(a) the CAE network used for the pre-training of the unLRG sample and 
(b) the neural network for the fine-tuning of the LRG sample. The convolutional layers are marked in blue, and the fully connected (FC) layer is marked in green. The image size (black), kernel size (black) and number (gray) of each convolutional layer and the length (red) of the FC layer are also marked.}
\end{figure}

The schematic architecture of the autoencoder is shown in Figure~\ref{fig.cae}a, which basically consists of several convolutional layers assigned with multiple $3\times3$ pixel kernels according to state-of-the-art neural networks theories (e.g., AlexNet and VGG; \citealt{Krizhevsky2012,Simonyan2014,Ma2017cavity}). The encoder block functions via a feature extractor, which focuses on local image features from lower to higher orders through successive convolution operations performed in multidimensional (multichannel) convolutional layers and employs a classifier consisting of one fully connected layer to transform the derived feature maps into feature vectors to implement the classification task (see Fig.~\ref{fig.cnn}b). 
 
The decoder block is a mirror of the encoder block. It is constructed by reversing the positions of the convolutional and fully connected layers in the encoder, and the layers in the decoder share the same structures of weights as their mirror counterparts in the encoder. The image reconstruction function of the decoder not only assists the unsupervised learning and training of the network parameters through analyzing the residuals of the model but also acts as a generator of radio AGN images when fed with proper feature vectors sampled from the probability distribution of the extracted features by the encoder. Similar architectures have been widely applied in a variety of areas, especially when simulating or generating data is important, such as in bottleneck training in speech recognition~(e.g., \citealt{Yu2014}), in the optimization of discriminator and generator in image simulation~(e.g., \citealt{Goodfellow2014GAN}), and in artistic styles learning~(e.g., \citealt{Radford2015}).

Based on a series of tests, we embed five convolutional layers into the encoder block (Fig.~\ref{fig.cae}a) to hierarchically extract image features, so that the sizes of the input images reduce reasonably from the original 140 $\times$ 140 pixels  in the first layer to 5 $\times$ 5 pixels in the fifth layer. After the fifth convolutional layer we use one fully connected layer, which is named the encode layer and is used to conduct the classifications, to further compress the derived features into a 64-dimension vector; including more fully connected layers will not significantly increase the performance of classification in this case (see Section~\ref{sec.FetLength} for a discussion). Immediately after each convolutional layer we apply a rectified linear unit (ReLU) to generate the activation function, which is nonsaturating and can help increase the convergence speed by truncating negative output into zero~\citep{Glorot2011}. The number of kernels (i.e., the number of channels) of each convolutional layer and the corresponding kernel sizes are listed in Table~\ref{tab.cae}. 

Because the decoder block is designed as the reverse of the encoder block, we feed the output of the encoder (i.e., the extracted feature vector) into the decoder and compare the output of the latter, i.e., the reconstructed images, with the original images using the mean squared error to evaluate the performance of the autoencoder. Obviously, such an autoencoder can also be used to generate radio AGN images as a simulator by feeding the randomly obtained 64-length feature vectors sampled from the distributions of the extracted morphological features by the encoder into the decoder block (see, e.g., our recent work in \citealt{Ma2018RCAE} as an example).
 \subsection{Training Strategy\label{sec.TrnStg}}
\begin{figure}[]
\centering
\includegraphics[width=0.48\textwidth]{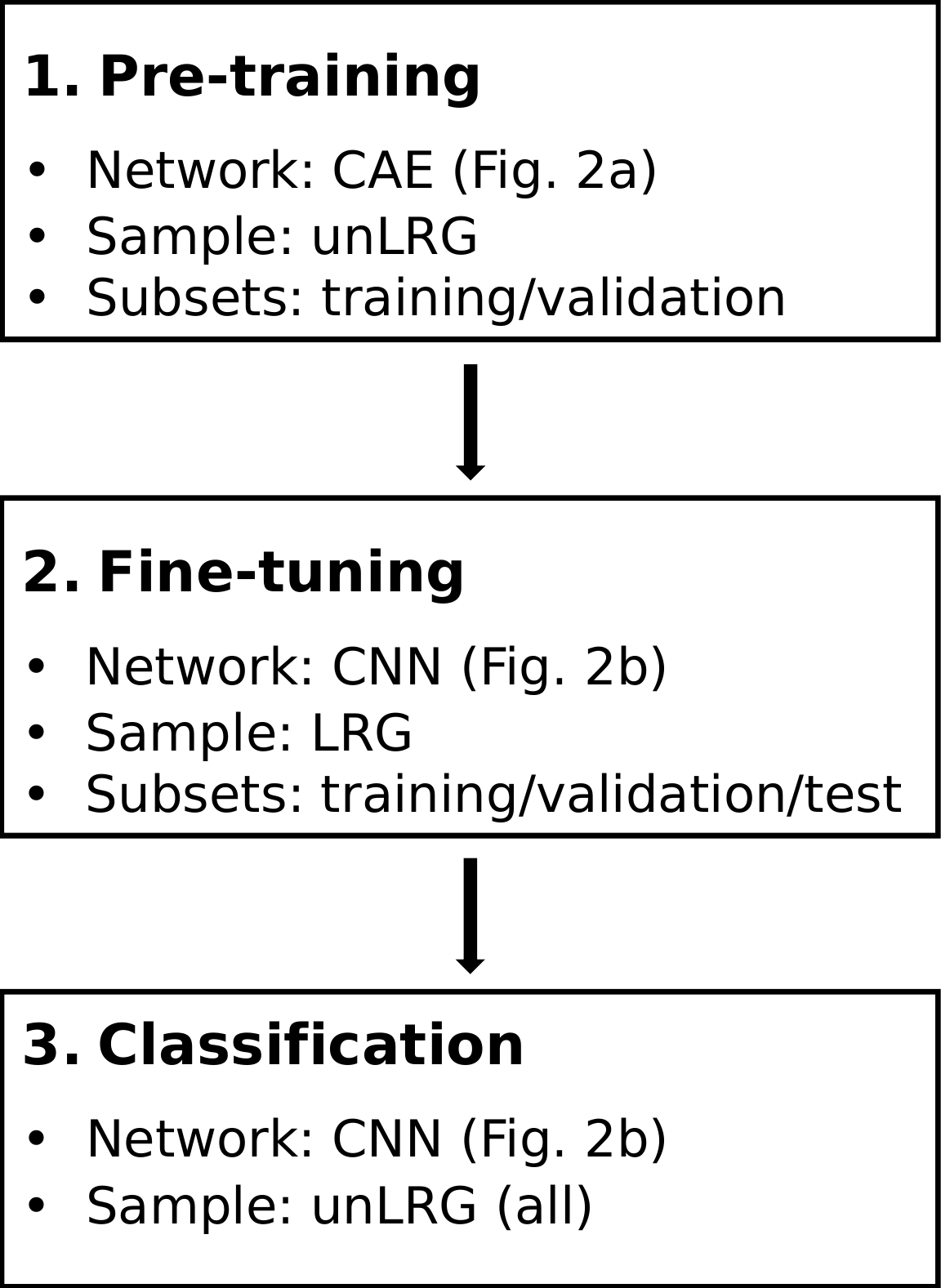}
\caption{Framework of the PFC three-step training strategy.\label{fig.pfc}}
\end{figure}

Given that the size of the labeled sample is relatively small, we apply an effective training strategy that integrates the ``pre-training" with the ``fine-tuning" process~\citep{Hinton2012,Girshick2014,Christodoulidis2016} by adopting the following steps (Fig.~\ref{fig.pfc}): (1) Pre-training---Apply the self-taught autoencoder to the unLRG sample (14,245 unlabeled radio AGNs) to pre-train a large number of weight and bias parameters of the deep CNN model, in order to deeply learn the representations of the radio AGN images and to avoid as much as possible the overfitting problem usually associated with the learning on a small sample. (2) Fine-tuning---Append a softmax layer (e.g., ~\citealt{Hinton2012}) to the end of the encoder that has been pre-trained in the first step, and use it to classify the LRG sample (1442 radio AGNs) with an emphasis on the training of the fully connected layer, which is used to carry out classification. (3) Classification---Apply the fine-tuned encoder to classify the unLRG sample.

Our approach of training the neural network with the above three steps (PFC steps hereafter) has several advantages. First, when there is a lack of sufficiently large labeled samples (a sample containing $\gtrsim 10^{4}$ labeled radio AGNs may be necessary in our case, given the level of image complexity; see a similar case in \citealt{LeCun2015} and the theoretical discussion in \citealt{Esteva2017}) for the training of a complex neural network, a straightforward training with small labeled samples does not help lower the risk of model overfitting and poor generalization capability, i.e., low predictive precision on new images. For example, the neural network is likely to overreact to minor, unimportant fluctuations in the training data and learn inessential random features. This problem is overcome by applying the PFC steps. Second, our approach enables one to perform the classification of a sample of about 15,000 radio galaxies efficiently and accurately in several seconds even on a portable computer equipped with a single GPU of moderate memory ($\lesssim$10 GB), not just on a workstation with multiple GPUs ($\gtrsim$100 GB), which other approaches usually require. This is because the resource consumption of the deep neural network in the encoder block of our autoencoder is significantly lower than that of other complex networks, e.g., the GoogLeNet~\citep{Szegedy2015,Aniyan2017}. This allows astronomers to inspect a newly obtained sample conveniently. In addition, our autoencoder can be used as a random-sample generator for morphology-based studies of radio galaxies. 
 \subsection{Convolution and Deconvolution}
\subsubsection{Convolution\label{sec.conv}}
A convolution operation is usually a linear transformation that multiplies an input image with a kernel that slides across the input image to produce a matrix called an output feature map~\citep{Dumoulin2016}. For a CNN containing a set of $L$ layers, each convolutional layer is assigned a set of kernels $W^{l}_{n}$ and biases $b^{l}$ to form $N^l_{\rm ch}$ channels, where $l =1,~2,~\dots,~L$ and $n=1,~2,~\dots,~N^{l}_{\rm ch}$ denote the ordinal numbers of the layers and channels, respectively. Thus for the $n$th channel in the $l$th layer of the network, the output feature map $O^{l}_{n}$ can be written as  
\begin{equation}
\label{eq.conv}
O^{l}_{n} = 
\begin{dcases}
W^l_{n} * I_0 + b^l_{n}, &{\rm if}~l = 1, \\
\sum^{N^{l-1}_{\rm ch}}_{m=1}{W^{l}_{n} * I^{l}_{m}} + b^{l}_{n}, &{\rm if}~l =2,~3,~\dots,~L,
\end{dcases}
\end{equation}
where $I_0$ is the original image fed into the network, $I^{l}_{m} \equiv O^{l-1}_m$ ($l\geq2$) is the $m$th input image ($m = 1,~2,~\dots,~N^{l-1}_{\rm ch}$), the notation `$*$' is the convolution operator, and $b^{l}_{n}$ is the bias.

Once an input is given, the corresponding output of a convolutional layer is uniquely determined by its kernels and biases and the choices of stride and zero-padding modes. For simplicity, in this work we always scale the images, as well as the kernels, as square matrices. The initial values assigned to the elements in the kernels are randomly generated with truncated normal Gaussian distribution, and corresponding biases are initialized with constants~(\citealt{TensorFlow2015-whitepaper}). We select half-padding (i.e., same padding) $p = \lfloor{k/2}\rfloor$, where $k$ is the size of the squared kernel, to keep the shape and set the size of the input. We also set the stride, the step length to slide the kernel, to $s = 2$, which leads to a linear pooling (i.e., subsampling or downsampling), to decrease the number of free parameters and meanwhile extract the most remarkable features. 
\subsubsection{Deconvolution (Transposed Convolution)}
The decoder block of the autoencoder can be regarded as a mirror of the encoder block, which can be derived by reversing the inputs and outputs of the CNN described in \S\ref{sec.conv} in turn. Similar to the convolution operation (Eq.~\ref{eq.conv}), the deconvolution can be expressed as 
\begin{equation} 
\label{eq.deconv}
\begin{dcases}
O^{l'}_{n'} =\sum^{N^{l'}_{\rm ch}}_{m=1}{[W^{l'}_{m}]^{T} *^{T} I^{l'}_{m}} + b^{l'}_{n'}, &{\rm if}~l' =1,~2,~\dots,L-1, \\
O^{l'} = \sum^{N^{l'}_{\rm ch}}_{m=1}{[W^{l'}_{m}]^{T} *^{T} I^{l'}_{m}} + b^{l'}, &{\rm if}~l' =L,
\end{dcases}
\end{equation}
for the $n'$th kernel channel ($n'=1,~2,~\dots,~N^{L-l'}_{\rm ch}$, and $n'\equiv1$ when $l'=L$) in the $l'$th layer, where $I^{1}_{m}$ ($ m=1,~2,~\dots,~N^L_{\rm ch}$) denotes the images of the first layer in the decoder block (i.e., the feature maps recovered from the $L$th encode layer), $I^{l'}_{m} \equiv O^{l'-1}_{m}$ ($l'\geq2$) is the $m$th input image of layer $l'$ ($m = 1,~,2~,\dots,~N^{L-l'}_{\rm ch}$), $b^{l'}_{n'}$ is the bias, and the superscript $T$ denotes the transpose arithmetic~\citep{Dumoulin2016}. A detailed explanation of the transpose convolution is illustrated in Appendix~\ref{sec.TransConv} with examples.

\subsection{Feature Visualization with $t$-SNE\label{sec.tsne}}
In order to evaluate the performance of our autoencoder, we visualize the classification of the LRG sample, which is achieved by analyzing the multidimensional feature vectors, in two-dimensional space (the visualization space) by applying the $t$-SNE algorithm, a dimension reduction algorithm that can preserve the similarities between vector points (i.e., distances) when mapping data from high-dimensional space onto lower dimensions~(e.g., \citealt{Laurens2008,Esteva2017}). 

To be specific, we use a set of multi-dimensional vectors $\{\bf{x_{1}},~\bf{x_{2}},~\dots,~\bf{x_{N_{\rm LRG}}}\}$, where $N_{\rm LRG}$ is the number of the LRG sample. These vectors form a distribution $\mathcal{P}$, in which $p_{ij}$ ($i,~j = 1,~2,~\dots,~N_{\rm LRG}$) denotes the probability of similarity between any two vectors $\bf{x_i}$ and $\bf{x_j}$. If we further use $p_{j|i}$ to define the conditional probability of $\bf{x_j}$ when $\bf{x_i}$ is given, by assuming the Gaussian distribution, we have 
 \begin{equation}
p_{j|i} = \frac{\exp{(-\lVert{\bf{x_{i}} - \bf{x_{j}}}\rVert}^2 /
2\sigma^2_{i})}{\sum_{k\neq i }{\exp{(-\lVert{\bf{x_{i}} -
\bf{x_{k}}}\rVert}^2 / 2\sigma^2_{i})}},
\end{equation}
where $\sigma^2_{i}$ is the variance of $\bf{x_i}$ (see~\citealt{Hinton2002SNE} for details), and  
\begin{equation}
p_{ij} = \frac{p_{j|i} + p_{i|j}} {2N_{\rm LRG}}
\end{equation}
by symmetry. 

Similarly, in the visualization space we use a vector set $\{\bf{y_{1}},~\bf{y_{2}},~\dots,~\bf{y_{N_{\rm LRG}}}\}$ to describe the distribution $\mathcal{Q}$ for the classified radio AGNs and $q_{ij}$ to represent the probabilities of similarities between any two vectors. As shown by \citet{Laurens2008}, $q_{ij}$ appropriately preserves the similarities between the vectors in the high-dimensional space, i.e., $p_{ij}$. Unlike the case of the distribution $\mathcal{P}$, for which the Gaussian distribution is assumed, \citet{Laurens2008} employed a long-tailed Student's $t$-distribution with one degree of freedom to describe $\mathcal{Q}$ to avoid abnormal points, which is written as
\begin{equation}
q_{ij} = \frac{(1 + {\lVert {{\bf y_{i}} - {\bf y_{j}}}
\rVert}^2)^{-1}}{\sum_{k\neq m}{(1 + {\lVert {{\bf y_{k}} - {\bf y_{m}}}
\rVert}^2)^{-1}}}.
\end{equation} 
By applying the Kullbach--Leibler (KL) divergence~\citep{Kullback1951}  
\begin{equation}
\label{eq.kl}
\mathrm{KL}(\mathcal{P}||\mathcal{Q}) = \sum_{i\neq j}{p_{ij}\log{\frac{p_{ij}}{q_{ij}}}},
\end{equation}
as the objective function, we can evaluate the similarity between distributions $\mathcal{P}$ and $\mathcal{Q}$, and locate ${\bf y_i}$ by minimizing Equation~(\ref{eq.kl}) with respect to ${\bf y_i}$ with 
\begin{equation}
\frac{\partial{\mathrm{KL}}}{\partial{\bf{y_i}}} =
4\sum_{j}{(p_{ij}-q_{ij})({\bf y_i}-{\bf y_j}) (1+(\lVert{{\bf y_i} -{\bf y_j}}\rVert^{2})^{-1}}.
\end{equation}
 
\section{Application of MCRGNet to Radio AGN Samples\label{sec.result}}
\subsection{Preparation of the Input Images\label{sec.pre}}
We apply an approach that consists of the following three steps to prepare both the labeled and unlabeled images before we feed them into the autoencoder. First, we define a region of interest (ROI) in each image by cropping a box region with a size of 140 $\times$ 140 pixel ($4'.2 \times 4'.2$) from the original 150 $\times$ 150 pixel ($4'.5 \times 4'.5$) image. Next, in order to improve the contrast of the ROI of each image, we carry out sigma clipping to limit the image dynamic range by setting the parameter $\sigma$ to 3 and the number of iterations to 50, and discard pixels dominated by the background noise. Finally, in order to further avoid overfitting of the network, we perform the augmentation operation to increase the number of images in both the LRG and unLRG samples by applying the following operations to each image:
\begin{enumerate}
\item
Randomly flip the sigma-clipped image. The flipping mode is either left-to-right, or top-to-bottom, or diagonal;
\item
Rotate the flipped image with an angle $\theta$ generated randomly between $0^{\circ}$ and $360^{\circ}$.
\end{enumerate} 

\begin{figure}[]
\centering
\includegraphics[width=0.85\textwidth]{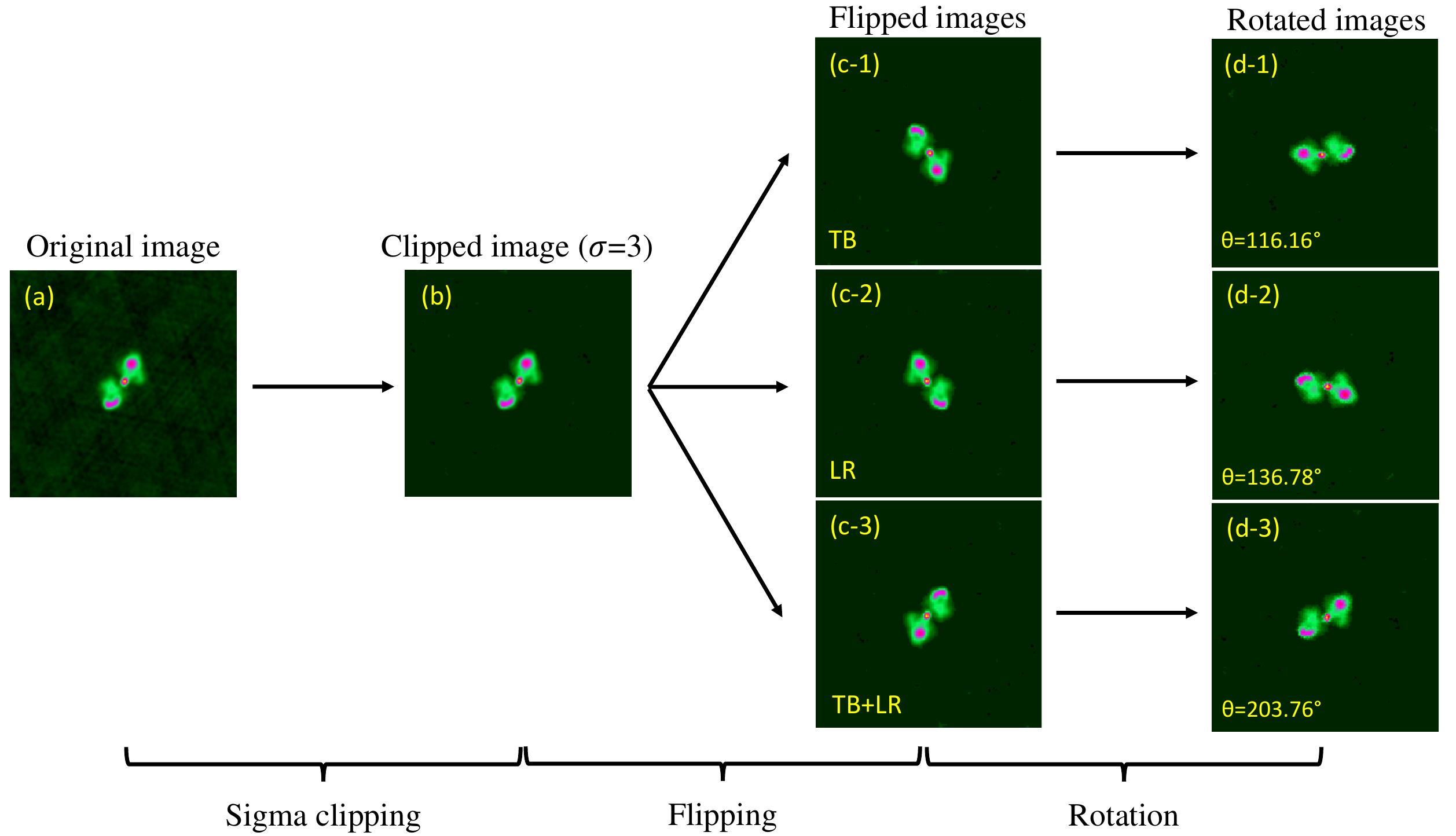}
\caption{An example from the augmented radio galaxies, i.e., J$145941.95+290331.8$ (XRG), which is located at $z$ = 0.146. After sigma clipping, the image is randomly flipped either top-to-bottom (TB), or left-to-right (LR), or both ($TB+LR$), and then randomly rotated with an angle in $[0, \pi)$.\label{fig.clip_aug}}
\end{figure}

For each unLRG image, the number of augmentation operations is 10, which results in a total of around $1.43\times10^{5}$ images for the pre-training. For each LRG image, however, the number of augmentation operations is determined according to the morphological type of the radio AGN, in order to balance the numbers of different morphological types in the dichotomous classification tree. Since on the classification tree (Fig.~\ref{fig.tree}) the two sub-branches of any branch should have nearly the same AGN numbers, the ratios of the numbers of the compact, FRI, FRII, BT, XRG, and RRG types are set to 8:2:2:2:1:1 after augmentation operation, which results in about $3.9\times10^4$ images in total (38,882, 19,554, 9731, 9823, 4900, and 4923 images for the binary classification branches compact$--$extended, typical$--$atypical FR, typical FRI$--$FRII, BT$--$irregular, FRI-like$--$FRII-like, and XRG$--$RRG, respectively; Table~\ref{tab.aug}). One example of the flipping and augmentation operations is displayed in Figure~\ref{fig.clip_aug}.

The augmented unLRG sample is then divided into independent training and validation subsets, which are used to optimize the parameters of the neural network and to evaluate the learning performance (i.e., the ability to reconstruct the images from which features are learned) to avoid the overfitting risk for each epoch, respectively. The training and validation subsets are divided at the ratio of $4:1$, therefore the two subsets have about $1.14\times10^5$ and $2.85\times10^4$ augmented images, respectively. 

In the original LRG sample, the numbers of fine-tuning images (i.e., training and validation images) and testing images are set to 1154 (80\%) and 288 (20\%), respectively. Data augmentation is  performed only on the images used for fine-tuning, which are divided with a ratio of $4:1$ to obtain 31,106 training images and 7776 validation images (Table~\ref{tab.aug}). Note that the subset of testing images is applied for evaluating the classification performance (i.e., the ability to classify radio AGNs correctly) of the trained network; therefore it contains only the original LRG images that are not augmented.

\subsection{Training and Evaluation of the Neural Network\label{sec.TrEval}}
Following the strategy described in \S\ref{sec.TrnStg}, we first feed the unLRG training and validation subsets into the autoencoder to pre-train the weight and bias parameters, and then input the LRG training, validation, and test subsets into the neural network, which is complemented with a softmax layer at the end of the encoder block, to fine-tune the network parameters with an emphasis on the fully connected encode layer.
\subsubsection{Approach to Training and Evaluation\label{sec.AppTrnEva}}
In this work we pre-train and fine-tune the neural networks using the strategy of batch iteration~\citep{TensorFlow2015-whitepaper}. Instead of the whole training dataset, we separate it into batches (100 images for each) and input them into the network batch by batch to adjust the network parameters iteratively. One round of training with all the batches is called an epoch.

For the pre-training stage the augmented images in the unLRG training or validation subset are fed into the encoder block in batches to generate the 64-dimension feature vectors, which are successively fed as the input of the decoder block to reconstruct the images. We compare the reconstructed images with the corresponding original input images by calculating the mean squared residual error between them as the batch loss of pre-training, which is
\begin{equation}
\label{eq.rp}
R_{\rm{P}} = \frac{1}{N_{\rm{B}}}\sum^{N_{\rm{B}}}_{i=1}\sum^{N_{\rm{row}}}_{j=1}\sum^{N_{\rm{col}}}_{k=1}{|I^{\rm CAE}_{i,j,k} - O^{\rm CAE}_{i,j,k}|^2},
\end{equation}
where $N_{\rm B}$ is the number of images in one training batch, $N_{\rm{row}} = N_{\rm{col}}=140$ are the number of rows and columns of the images, $I^{\rm CAE}$ represents the input images of the encoder block, and $O^{\rm CAE}$ represents the reconstructed images of the decoder block. According to \citet{Goodfellow2016}, the loss of each batch is back-propagated to the network along the gradient decreasing direction to adjust network parameters until the loss converges to a constant value. This is achieved after about 500 epochs of pre-training are completed.

For the fine-tuning stage, a softmax layer is appended to the end of the encoder block, and the batch loss of fine-tuning $R_{\rm F}$,  is defined by averaging the cross-entropy of the augmented LRG images~\citep{Goodfellow2016} as 
\begin{equation}
R_{\rm{F}} = -\frac{1}{N_{\rm{B}}}\sum^{N_{\rm{B}}}_{i=1}\sum^{N_c}_{j=1}{y_{i,j}\cdot\log_2{p^{S}_{i,j}}},
\end{equation}
where $N_c = 2$ is the number of subbranches for each branch on the dichotomous classification tree and $y_{i,j}$, ($i = 1,~2,~\dots,~N_{\rm B}$ and $j=1,~2,~\dots,~N_c$; $y_{i, j} = 1$, if the source belongs to type $j$, and $y_{i,j} = 0$ if otherwise) is the one-hot probability of the source. $p^{S}_{i, j}$ represents an output of the softmax layer, which is the normalized probability of the source being classified as type $j$ in a certain batch. The batch loss converges to a constant after about 200 epochs of the fine-tuning are completed.

In the pre-training and fine-tuning stages we have employed the following standard rules.
\begin{enumerate}
\item
Dropout of weights with values approaching to zero ~\citep{Srivastava2014}, which helps avoid overfitting, is performed each time after the calculation in a convolutional or a fully connected layer is completed. The dropout rate is fixed at 50\% for each layer.
\item
In the pre-training stage the number of epochs is set to 500, which is determined by the convergence of training curves. A high initial learning rate of 0.0005 together with a decay rate of 0.95 to form the adaptive learning rate (\citealt{TensorFlow2015-whitepaper,Zeiler2012}) and an adaptive moment optimization function (ADAM;~\citealt{Kingma2014}) are both adopted. In the fine-tuning stage the number of epochs of each sub-classifier is set to 200 also according to the convergence of the training curves, and a lower initial learning rate of 0.0001 is adopted to carefully fine-tune the parameters.
\item
In either the pre-training or the fine-tuning stage, the corresponding validation loss $R_{\rm P}$ or $R_{\rm F}$, which is obtained by feeding the unLRG or the LRG validation subset into the autoencoder or CNN, is used to evaluate the over-fitting risks for each epoch. If the validation loss does not descend or even increase while the training loss decreases steadily to convergence, the network parameters may be overfitted. Note that the validation loss is not used for parameters optimizations.
\end{enumerate}

\subsubsection{Results of Training and Evaluation\label{sec.TraAndEval}}
\begin{figure}[]
\centering
\includegraphics[height=0.5\textwidth]{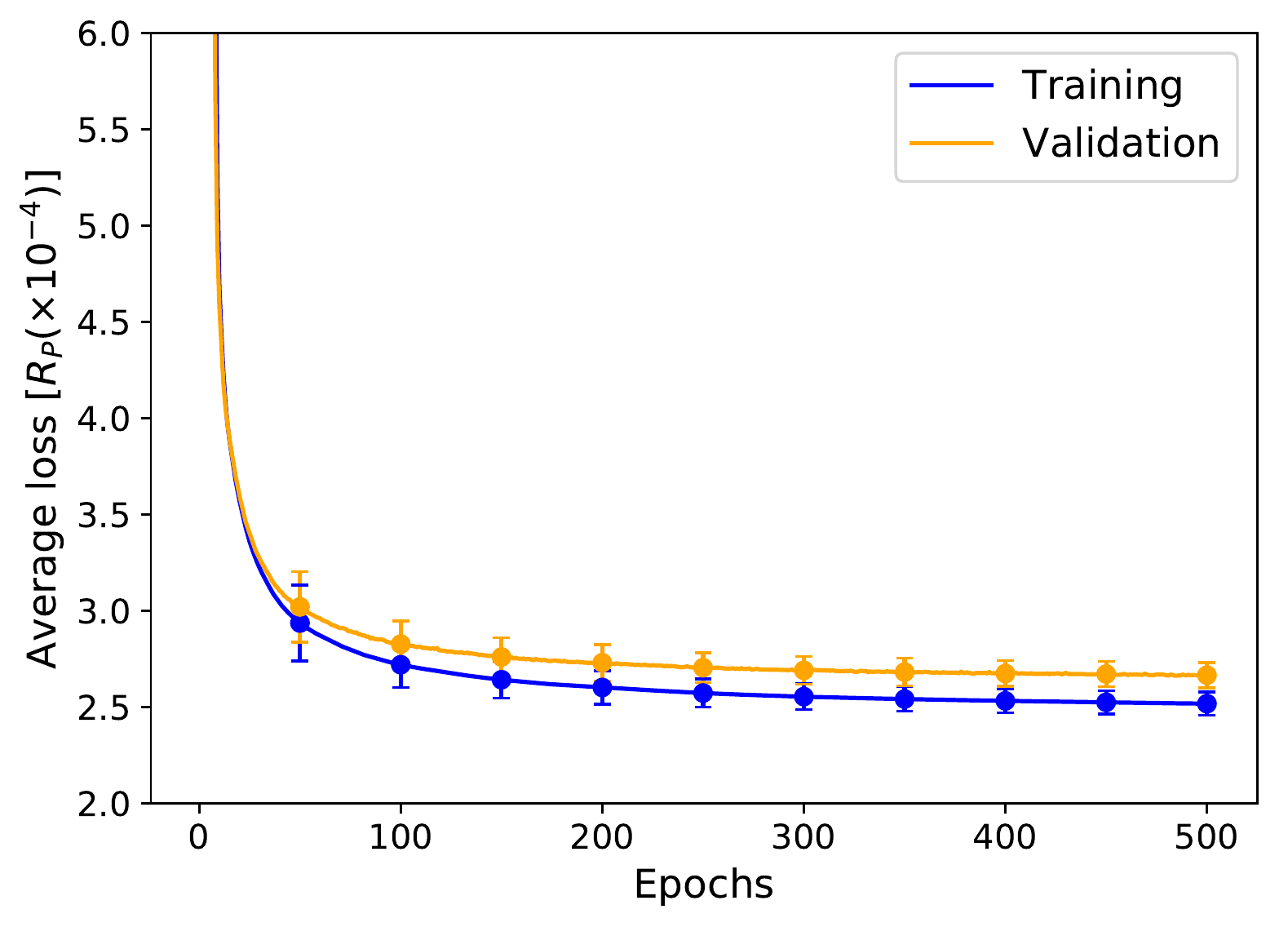}
\caption{Average loss $R_P$ calculated for the pre-training of the autoencoder network.\label{fig.lossp}}
\end{figure}

\begin{figure}[]
\centering
\includegraphics[height=0.84\textheight, width=0.75\textwidth]{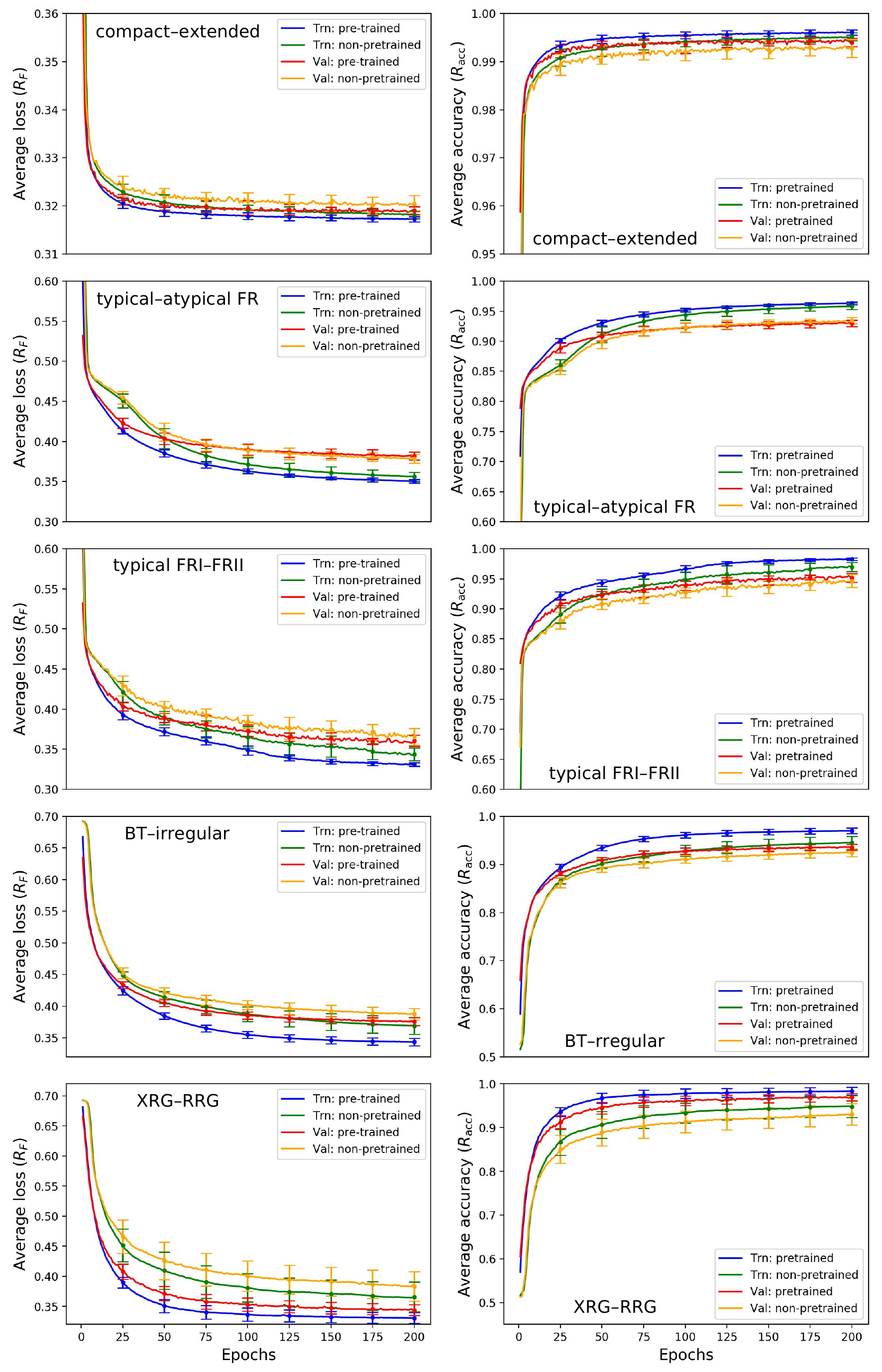}
\caption{Average loss $R_{F}$ (left) and accuracy $R_{\rm acc}$ (right) calculated for five dichotomous classification branches, i.e., compact--extended, typical--atypical FR, typical FRI--FRII, BT--irregular, and XRG--RRG. Results for both training and validation subsets are presented with or without pre-training (\S\ref{sec.pre}). \label{fig.loss_acc_cmp}}
\end{figure}

In Figures~\ref{fig.lossp} and ~\ref{fig.loss_acc_cmp}, we illustrate the training and validation loss curves obtained in the pre-training stage and fine-tuning stage, respectively. For comparison, in the fine-tuning stage we also plot the loss curves obtained when the pre-training process is switched off for comparison as well as the classification accuracy $R_{\rm acc}$, which is defined as the ratio of the number of correctly classified sources to the actual source number (see also Eq.~\ref{eq.acc}). To evaluate the robustness of the network, each training process has been conducted for 10 times, and the means and 68\% confidence limits of the losses and accuracies are calculated for every 25 epochs (see Fig.~\ref{fig.loss_acc_cmp}). In either the pre-training stage or the fine-tuning of all the subclassifiers, both the training loss and the validation loss decrease monotonically, and the training loss becomes lower than the corresponding validation loss after about 10 epochs, and this remains unchanged until the pre-training or fine-tuning is completed. These indicate that the trained parameters in the neural network are not overfitted. 

In the fine-tuning stage, except for the typical$--$atypical FR branch, the average fine-tuning loss $R_\mathrm{F}$ decreases significantly faster, and the neural network achieves better classification accuracies after pre-training is applied.  As for the typical$--$atypical FR branch, because (1) the sources therein are comparatively easier to classify due to their distinctive morphological features and (2) the number of images in the training subset is 19,554, the pre-training becomes more or less insignificant.

In order to further evaluate the performance of the neural network, we have employed three indices, i.e., sensitivity $R^t_\mathrm{sen}$, accuracy $R^t_\mathrm{acc}$, and precision $R^t_\mathrm{pre}$, to describe the classifier performance of the results for LRG samples~\citep{Khalaf2016,Ma2017cavity}. They are defined as follows:
\begin{align}
R^t_\mathrm{sen} &= N^t_\mathrm{TP} / (N^t_\mathrm{TP} + N^t_\mathrm{FN}),\label{eq.sen}\\
R^t_\mathrm{acc} &= (N^t_\mathrm{TP}+N^t_\mathrm{TN}) / N_\mathrm{LRG}, \label{eq.acc}\\
R^t_\mathrm{pre} &= N^t_\mathrm{TP} / (N^t_\mathrm{TP} + N^t_\mathrm{FP}), \label{eq.pre}
\end{align}
where the superscript $t$ denotes the morphological type ($t=1,~2,~\cdots,$ and 6, corresponding to compact, typical FRI, typical FRII, BT, XRG, and RRG, respectively), $N_\mathrm{LRG}$ represents the total number of radio AGNs in the original LRG sample, $N^t_\mathrm{TP}$ ($N^t_\mathrm{TN}$) is the number of type $t$ (non-type $t$) radio AGNs that are correctly classified (T), and $N^t_\mathrm{FN}$ ($N^t_\mathrm{FP}$) is the number of type $t$ (non-type $t$) radio AGNs that are misclassified (F). We list the calculated indices in Table~\ref{tab.eval} with 68\% error bars, and find that our neural network can achieve a satisfactory classification. The classifications of compact, typical FRI, typical FRII, and BT types are very accurate, although the classifications of XRG and RRG types are relatively poor, possibly due to insufficient training/learning caused by their small sample scale. 

\begin{figure}[]
\centering
\includegraphics[width=0.95\textwidth]{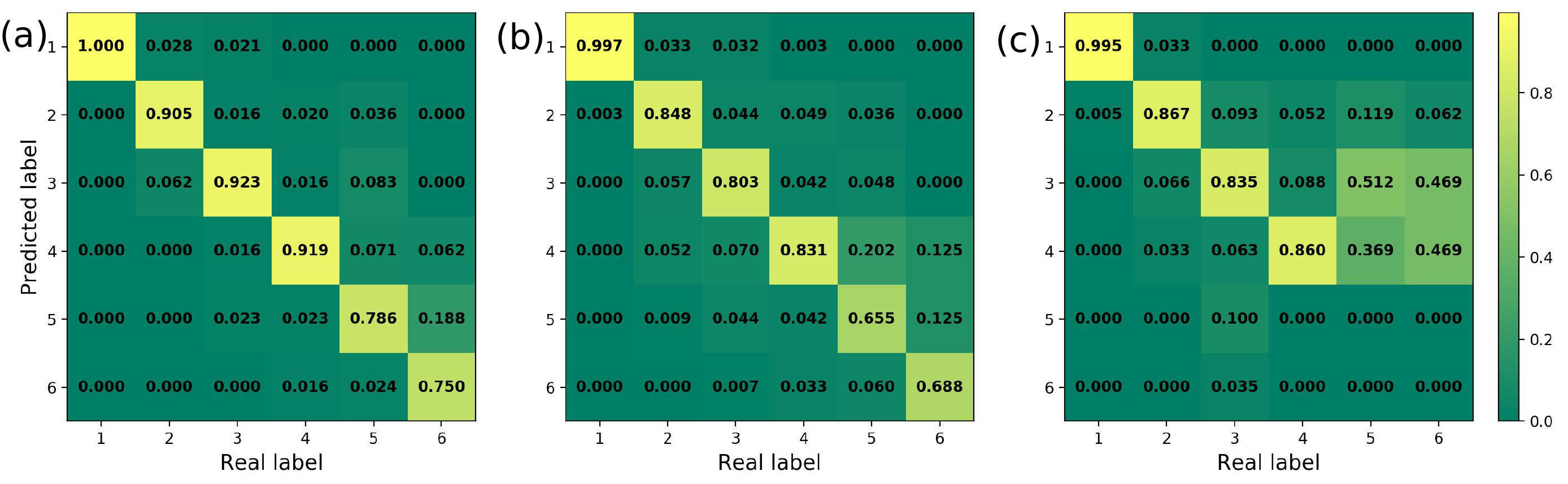}
\caption{Confusion matrices obtained for the LRG sample by training the CNN with (a) and without (b) pre-training using the tree strategy and (c) by applying the non-tree strategy. Labels from 1 to 6 are for compact, FRI, FRII, BT, XRG, and RRG sources, respectively. The numbers in each column are normalized. \label{fig.cmlrg}}
\end{figure}

\begin{figure}[]
\centering
\includegraphics[width=0.95\textwidth]{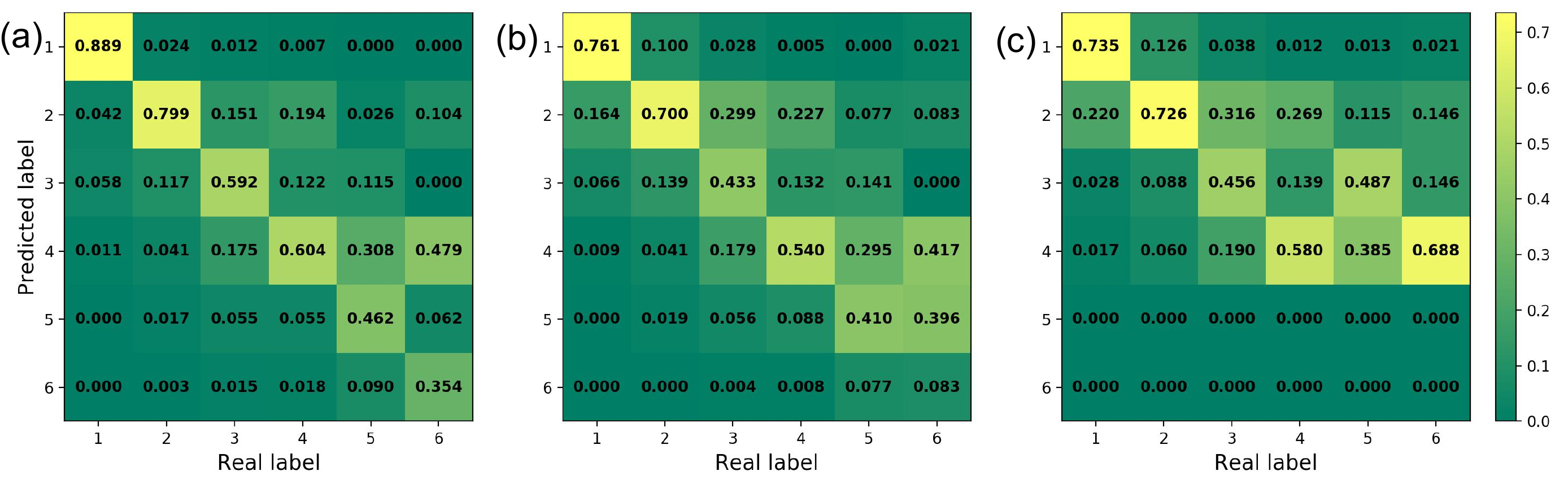}
\caption{Confusion matrices obtained for the unLRG sample by training the CNN with (a) and without (b) pre-training using the tree strategy and (c) by applying the non-tree strategy. Labels from 1 to 6 are for compact, FRI, FRII, BT, XRG, and RRG sources, respectively. The numbers in each column are normalized.\label{fig.cmulrg}}
\end{figure}

The results obtained by no pre-training strategy are also listed in Table~\ref{tab.eval} for a comparison. The averaged precision gain between pre-training and no pre-training is $7.77\% \pm 3.02\%$, which indicates that pre-training indeed improves the network performance. Besides, we show the confusion matrices of classifying the LRG sample with and without pre-training strategies in Figure~\ref{fig.cmlrg} for visualizing the difference between their performance. Though the gains in precision and sensitivity after applying pre-training on the AGN types with large source numbers (e.g., compact, FRI, and etc.) might be relatively insignificant, which are significant for the XRG and RRG sources of small numbers, especially for the sensitivity. The same conclusion could also be obtained in the classification of the unLRG sample (Fig.~\ref{fig.cmulrg}; see \S\ref{sec.unLRG} for details). 

We apply MCRGNet pre-trained by the proposed PFC strategy on the LRG sample and list the results including the corresponding 68\% errors, in Table~\ref{tab.eval}, along with the results obtained without applying the pre-training. The corresponding averaged gains of the indices obtained after carrying out pre-training for each subtype are also shown in the same table. Compared with the results of \citet{Aniyan2017}, who trained their network with a relatively small labeled sample (716 sources), our classification precisions of both typical FRI and FRII sources are higher (95\% vs. 91\% and 92\% vs. 75\%), although that of BT sources is lower (87\% vs. 95\%). \citet{Aniyan2017} also reported that the recall (i.e., the index $R_\mathrm{sen}$ in this paper), which reflected the ability to avoid misclassification due to the incorrect inclusion of the sources of other types, of the BT sources obtained with their neural network was 79\%. They argued that this was because some sources in the BT sources were misclassified as FRIIs, which was supported by the high recall but the low precision of FRII sources. As can be seen in Table~\ref{tab.eval}, the sensitivities $R_{\rm sen}$ for the BT and FRII sources derived by our approach are both $>91\%$. This, combined with the corresponding high precisions, suggests an improved classification of BT and typical FRII radio AGNs.  

\subsubsection{Feature Visualization of the LRG sample}
We show the results of the dimension-reduced feature visualization of the LRG sample classification with the $t$-SNE in Figure~\ref{fig.tsne}, in which the subsamples of different types are marked with different colors. We find that the points of each type gather into a cluster that overlaps only slightly with those of other types, except for the case between FRIIs and XRGs, and the case between compacts and FRIs. By carrying out visual inspection, we find that the FRII$--$XRG confusion is caused when the minor pair of jets in the XRG is dominated by the major pair, and the compact$--$FRI misclassification is caused when an FRI shows small-scale and weak jets.

To show the merit of using a dichotomous classification tree structure (Fig.~\ref{fig.tree}), we have  trained a CNN which shares the same structure as the encoder block illustrated in Figure~\ref{fig.cae}a but has a six-node softmax layer as a classifier and used it to directly classify the LRG images into the six types. As shown in Figure~\ref{fig.tsne}c the result is unacceptable, because the points of the different types overlap with one another severely. Based on this we may conclude that, compared with direct six-type (non-tree) classification, the application of a dichotomous tree leads to significantly better performance.

In Figure~\ref{fig.cmlrg} we show the $6\times6$ confusion matrices of the LRG classification obtained with and without pre-training and that obtained when direct six-type classification is performed. Compared to the dichotomous tree based methods, the direct training leads to the worst results, especially for XRG and RRG sources, whose numbers are relatively small. The same conclusion could also be found in Figure~\ref{fig.cmulrg} on the classification of the unLRG sample (see Section \ref{sec.unLRG} for details).

\subsection{Classification of the unLRG Sample\label{sec.unLRG}}
After deeply training the neural network with both unlabeled and labeled samples, we classify 14,245 radio AGNs in the unLRG sample and list the results in Tables~\ref{tab.unLRG} and \ref{tab.unLRG_vis}. In addition to the CNN classification, for each source we have also estimated the joint probability of the classification by cascadingly multiplying the probabilities of the nodes decided and given by the softmax layer at the corresponding branches (\citealt{Goodfellow2016}; see Appendix~\ref{sec.ProbBT} for details). Furthermore, if a source is classified as a BT, a subtype label (1 for FRI-like sources or 2 for FRII-like sources) is further presented, which is obtained by the CNN trained at the typical FRI$--$FRII branch.

In order to cross-check the results, we have also visually inspected all the unLRG images and classified the sample manually. We divide the images into three parts, and each part is inspected by at least three of the authors independently. If the three authors arrive at the same result, the visual classification is determined. Otherwise, an additional two authors will join, and the visual classification is determined when at least three authors agree on the result. The visual classification is also listed in Table~\ref{tab.unLRG}. 

We find that in general the CNN classification agrees with the human classification. About 80\% of the sources have the same CNN and manual labels, and this ratio is even higher for compact sources (89\%) and typical FRI sources (80\%). About 3\% of the small-scale FRIIs are misclassified by the neural network as FRIs, due to the small distances between the two lobes and their weak fluxes. About 7\% of the typical FRs are misclassified by the neural network as BTs, XRGs, and RRGs, because these FRs tend to have ambiguous morphologies and low PAPRs ($\leq 10.0$ dB).  In most cases, a low or relatively low ($\lesssim 0.90$) classification probability is found for the sources misclassified by the neural network. A further discussion on this issue will be presented in \S\ref{sec.DisunLRG}. 

\section{Discussion\label{sec.dis}}
\subsection{A Comparison between FRI and FRII Radio Galaxies\label{sec.cmpFR}}
\citet{Fanaroff1974} proposed a clear radio power transition between FRIs and FRIIs based on their study of the 3C (\citealt{Mackay1971}) radio sources. \citet{Ledlow1996} suggested that the transition was a function depending on optical luminosity and was most likely a result of the different lifetimes for sources with different radio powers. Subsequent studies (\citealt{Best2009} - 1083 FRIs and FRIIs; \citealt{Wing2011}  - 1,228 FRIs and 683 FRIIs; \citealt{Capetti2017b} - 122 FRIIs), however, do not support the claim that FRIs and FRIIs reside in distinctly different regions in the radio luminosity--optical magnitude plane ($L_{\rm radio} - M_{\rm opt}$ plane). With the large sample of radio AGNs classified in this work, which includes 5048 FRIs and 2083 FRIIs, it will be very interesting to investigate to what extent the distributions of FRIs and FRIIs differ in the $L_{\rm radio} --M_{\rm opt}$ plane. 

The 1.4 GHz radio luminosities of the sources in our sample are calculated with the flux densities measured at 1.4 GHz, as listed in Table~1 of \cite{Best2012}. The optical $r$-band luminosities are calculated by utilizing the photometric parameters and spectroscopic redshifts drawn from SDSS (\citealt{York2000}). As recommended by SDSS, we adopt the cmodel magnitude, one of the magnitude measurements provided in the SDSS database, to estimate the optical fluxes of radio AGNs. By using the SDSS photometric pipeline to combine the de Vaucouleurs and exponential models (into a linearly composite model or cmodel) to obtain best fits of the SDSS images, we obtain the composite flux of each source as follows:
\begin{equation}
F_{\rm composite} = f_{\rm deV}\times F_{\rm deV} + (1 - f_{\rm deV})\times F_{\rm exp},
\end{equation}
where $f_{\rm deV}$ is the coefficient of the de Vaucouleurs model; meanwhile $F_{\rm deV}$ and $F_{\rm exp}$ are the best-fit fluxes of the de Vaucouleurs and exponential models, respectively. With the apparent $r$-band magnitude $m_{r}$, which is calculated from $F_{\rm composite}$, we have the absolute $r$-band magnitude $M_{r}$ 
\begin{equation}\label{eq:absmag}
  M_{r}=m_{r} - 5\log\left( \frac{D_{\rm L}}{\rm pc}\right)+5 - A - K,
\end{equation}
where $D_L$ is the luminosity distance, $A$ is the correction for the Galactic extinction derived with the map of \citet{Schlegel1998}, and $K$ is the $k$-correction calculated by following~\citet{Blanton2007}.

\begin{figure}[]
\centering
\includegraphics[width=0.98\textwidth]{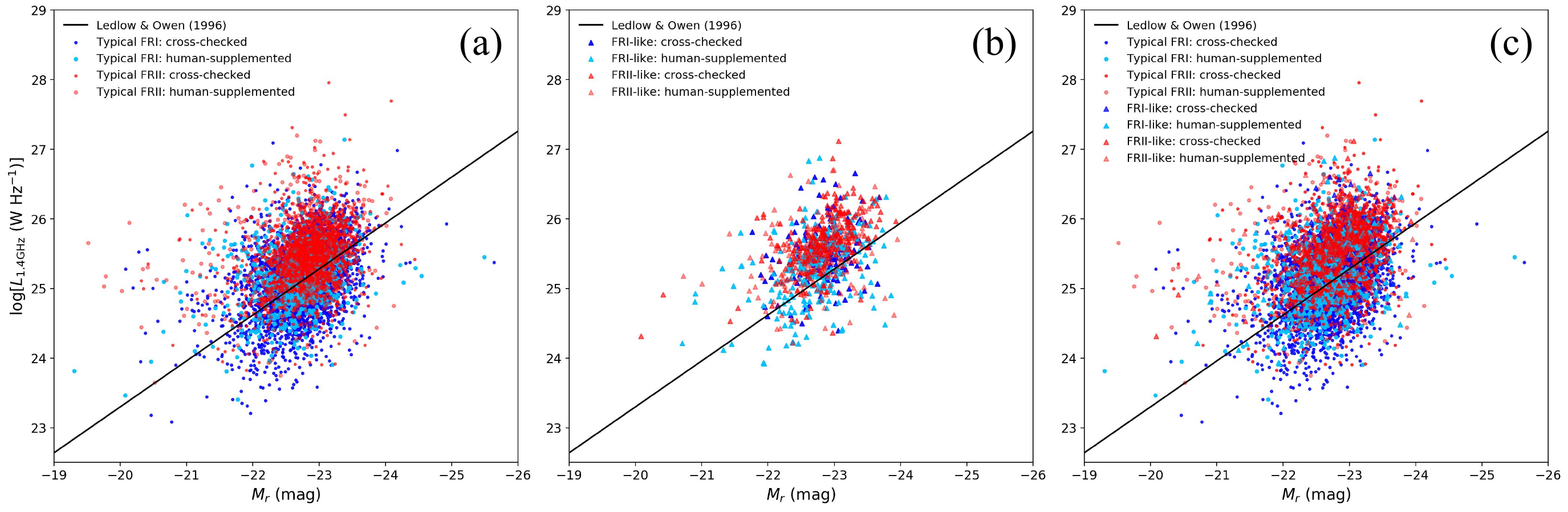}
\caption{Joint distributions of the unLRG FR samples between 1.4 GHz luminosities ($\log[L_\mathrm{1.4GHz}]$) and $r$-band absolute magnitudes ($M_r$). 
(a) Distributions of 4939 typical FRI sources (including 3953 sources cross-checked between the MCRGNet prediction and human-inspection, and the rest are only labeled by humans) and 2009 typical FRIIs (1547 sources cross-checked). (b) Same as (a) but for 420 FRI-like sources (143 sources cross-checked) and 440 FRII-like sources (238 sources cross-checked). (c) Combination of (a) and (b). The solid black line indicates the division of FRI--FRII reported by \citet{Ledlow1996}, to which corrections of magnitudes have been made by following \citet{Capetti2017a}.\label{fig.wlr}}
\end{figure}

In Figure~\ref{fig.wlr} we show the distributions of typical FRI (blue and cyan circles), typical FRII (red and coral circles), FRI-like (blue and cyan triangles), and FRII-like (red and coral triangles) sources in our sample on the $L_{\rm radio} - M_{\rm opt}$ plane, where the classification based on the cross-check between the MCRGNet prediction and human inspection is marked in blue and red and the classification purely made by humans (i.e., for sources not correctly classified by MCRGNet, which are termed ``human-supplemented") is marked in cyan and coral. For the cross-checked sources, a probability threshold of 0.90 is applied to filter those sources with smaller predicted probabilities (i.e., sources with high uncertainties; see Section \ref{sec.uncertainty} for a detailed discussion).

\begin{figure}[]
\centering
\includegraphics[width=0.95\textwidth]{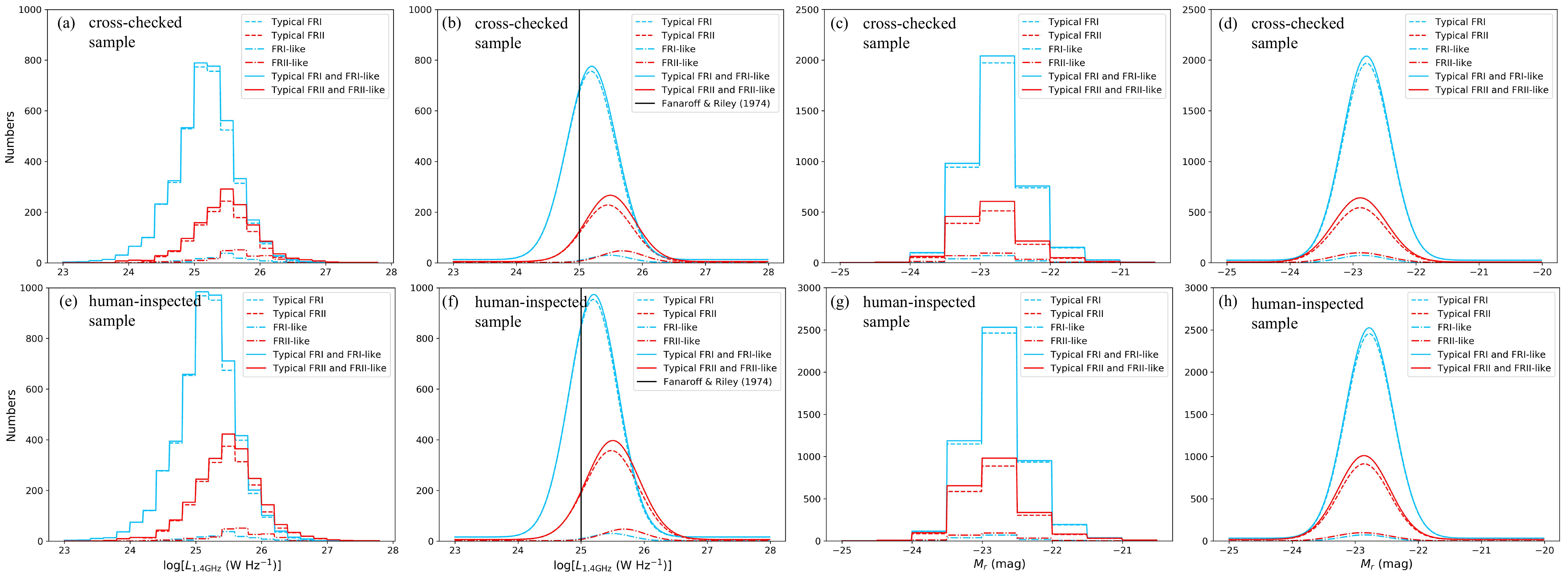}
\caption{Binned distributions and corresponding best Gaussian models of the unLRG FR samples as a function of the 1.4 GHz luminosity ($\log[L_\mathrm{1.4GHz}]$) or the $r$-band absolute magnitude ($M_r$). (a)-(d) are for the classification based on the cross-check between the MCRGNet prediction and human inspection, and (e)-(h) are for the classification based on the human-inspected classification. The black solid line indicates the division reported by~\citet{Fanaroff1974}.\label{fig.sdss}}
\end{figure}

The corresponding number distributions as a function of either $L_{\rm 1.4~GHz}$ or $M_{r}$, as well as their best Gaussian fits, are plotted in the lower panel of Figure~\ref{fig.sdss}  for a comparison. We find that at any given $M_{r}$ the distribution of typical FRIIs completely overlaps that of typical FRIs, although the latter extends to lower radio luminosities by about one order of magnitude. Meanwhile the FRI-likes and FRII-likes tend to share the same distribution. In addition to this, we also find that the division line between FRIs and FRIIs that was reported in \citet{Ledlow1996} crosses either of these distributions approximately in the middle. These results, which are obtained with a sufficiently large sample, confirm those reported in the previous studies \citep{Best2009,Lin2010,Wing2011,Capetti2017b} with good consistency, indicating that an abrupt separation between FRIs and FRIIs in the $L_{\rm radio}--M_{\rm opt}$ plane cannot be reproduced. A further discussion on this issue with an additional criterion on the samples will be presented in \S\ref{sec.uncertainty}. 

\begin{figure}[]
\centering
\includegraphics[width=0.98\textwidth]{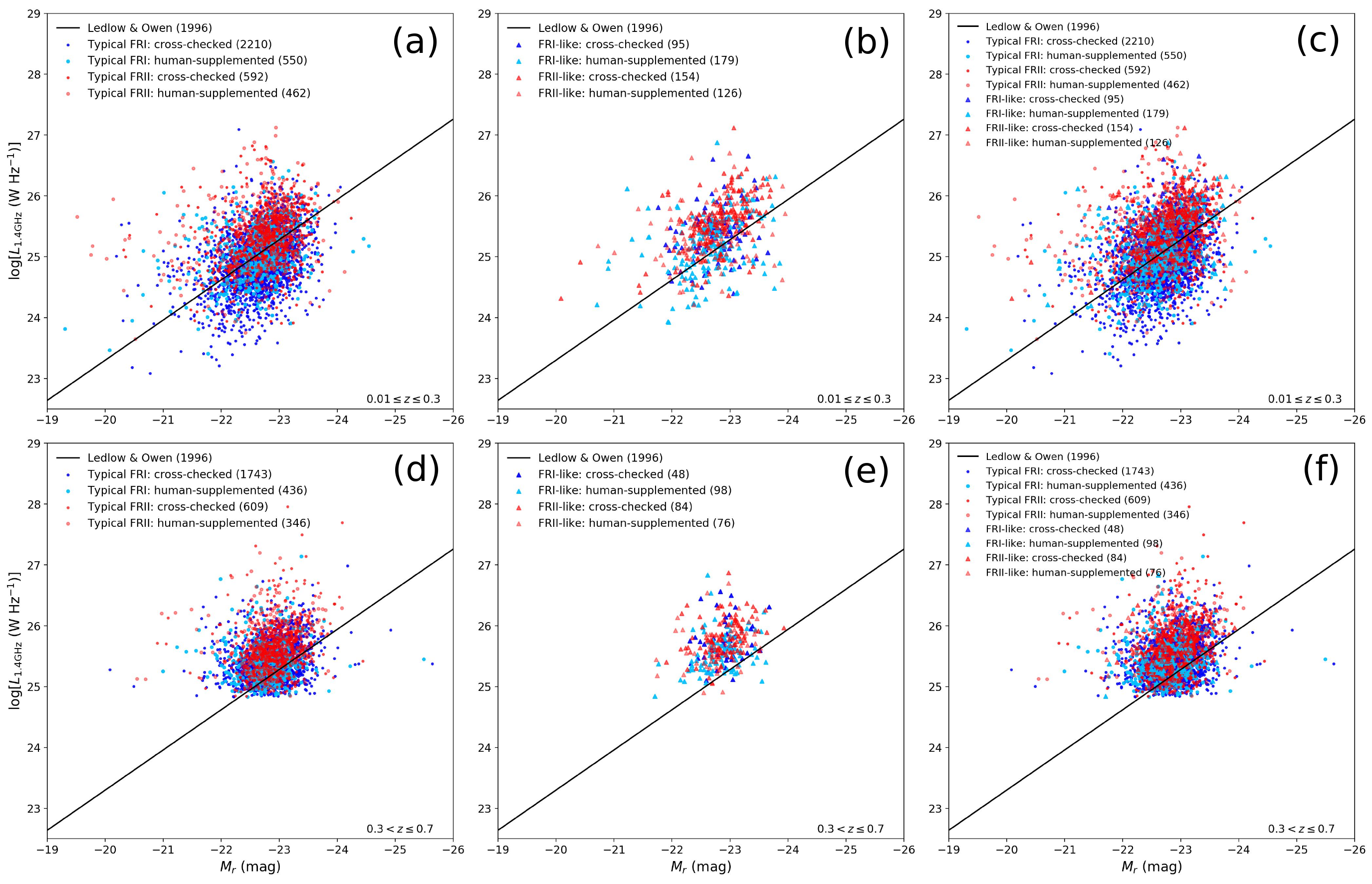}
\caption{Joint distribution of the unLRG FR samples between 1.4 GHz luminosities ($\log[L_\mathrm{1.4GHz}]$) and $r$-band absolute magnitudes ($M_r$), which are obtained in the redshift ranges $0.01 \leq z \leq 0.3$ and $0.3 < z \leq 0.7$ respectively. The symbols are the same as those used in Figure~\ref{fig.wlr}, and the corresponding source numbers are marked in the legends.\label{fig.zwlr}}
\end{figure}

In order to search for the possible redshift evolution of the $L_{\rm radio} - M_{\rm opt}$ distributions, we have divided the sources into $z \leq 0.3$ and $0.3 < z \leq 0.7$ subgroups and plot their distributions in Figure~\ref{fig.zwlr}, where the markers and colors of the sources are the same as in Figure~\ref{fig.wlr}. We find that roughly the same conclusion as above can be drawn for either of the two redshift ranges. Compared with FRIs located in $z \leq0.3$, FRIs in $0.3 < z \leq 0.7$ tend to be confined in relatively higher $L_{\rm 1.4~GHz}$ intervals, which is likely to be caused by the selection effect as fainter sources are hard to observe at higher redshifts and/or by the 5 mJy cut of NVSS to the sample. Therefore, we conclude that no redshift evolution is observed with our sample.

\subsection{A Comparison between Machine and Human Classification of the unLRG Sample\label{sec.DisunLRG}}
As can be seen in \S\ref{sec.TraAndEval}, the fully pre-trained and fine-tuned MCRGNet has achieved very satisfying classification precision on the LRG sample. Comparing the results obtained with MCRGNet and visual classification on the unLRG sample in \S\ref{sec.unLRG}, we find that the classification precisions for the compact and FRI sources are high ($\gtrsim 80\%$), but the precisions are low ($\lesssim 70\%$) for the FRII and BT sources and are actually worse for the XRGs and RRGs. As for the FRII and BT sources, we note that about 14\% of the former (308 FRIIs) and about 12\% of the latter (56 FRII-like BTs) are misclassified into each other by MCRGNet, which reproduces the result of \citet{Aniyan2017}. In addition, about 1\% of the FRIIs are misclassified into the compact category, and this can be attributed to relativistic beaming, by which a bright nearer lobe dominates its fainter counterpart in the images. As explained in \S\ref{sec.TraAndEval} for the LRG sample classification, the low precision for the XRG and RRG sources in the unLRG sample may also be due to their small numbers (78 XRGs and 48 RRGs) in the sample (14,245 sources) and their ambiguous morphologies.

\begin{figure}[]
\centering
\includegraphics[width=0.48\textwidth]{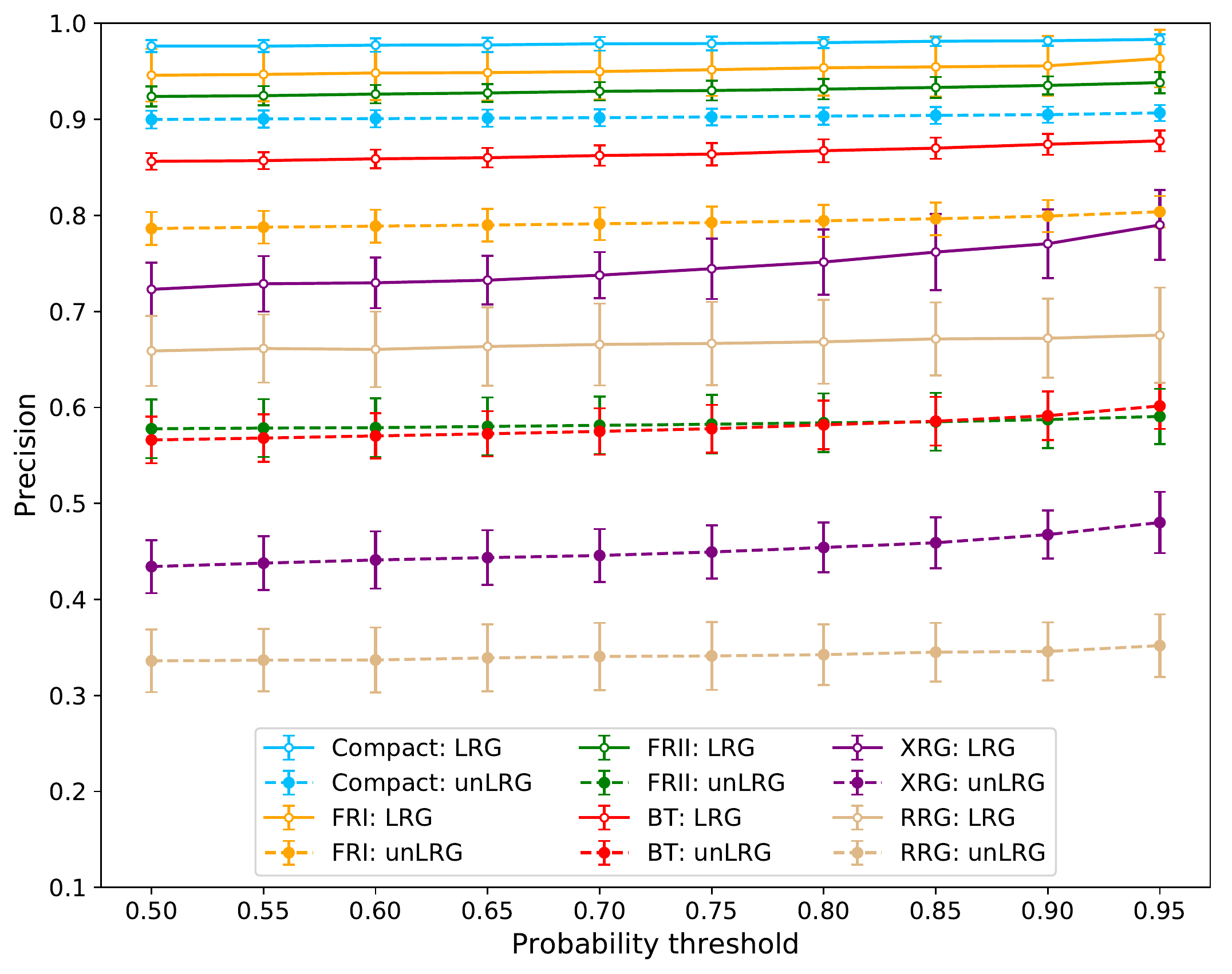}
\caption{Precisions of the proposed MCRGNet on the LRG and unLRG samples obtained by setting different probability thresholds.\label{fig.posthrs}}
\end{figure}

By cross-checking the misclassified sources, we find that most of them require more of us authors of us to join to vote them into certain AGN types. Also for these sources, most of them are found with low probabilities ($\lesssim0.90$) obtained by the CNN. Tentatively, we set a threshold to the probabilities of the final labels that are predicted by the dichotomous tree classifier, i.e., the objects with probabilities higher than the threshold are kept to calculate the precisions. The evolution of the precision of the classification for the six AGN subtypes is illustrated in Figure~\ref{fig.posthrs} as a function of the probability thresholds. We find that for the compact and FRI sources, the classification precisions are not nearly affected by changes in the probability threshold, which is consistent with the corresponding high probabilities ($\gtrsim 0.95$) of the labels predicted by our MCRGNet (see Table \ref{tab.unLRG}). However, for the XRGs and RRGs, which have relatively complex morphologies, the classification precision significantly increases as the threshold increases.

Actually, the problems described above are caused by the fact that the completeness of the labeled sample used to train the neural network is still inadequate. Although the pre-training on a large unlabeled sample (the unLRG sample) may be able to extract most of the features necessary to describe the radio galaxies, the fine-tuning process tends to enhance the features of existing instances carried in the LRG sample while weakening the uncarried. For example, no typical relativistically beamed FRII source is correctly involved in the LRG sample, as well as in other labeled FRII samples available in literature, which prohibits the recognition of the corresponding sources. This may be called a data-unseen problem, which could introduce uncertainty to the network (see Section \ref{sec.uncertainty} for a detailed discussion) and has been discussed in many machine learning based works, e.g., on image classification (e.g., \citealt{Chatzilari2015}) and object detection (e.g., \citealt{Zhu2016}). The solution to this problem is straightforward, i.e., enlarging the sample. Actually, providing the astronomers with a large morphologically classified sample which is as complete as possible is part of the motivation of our work. By repeating such a process to enlarge the labeled sample data scale, the data-unseen problem can be efficiently mitigated. This is called iteratively training thinking (e.g., \citealt{Ma2017cavity}). 

\subsection{The Use of One Fully Connected Layer and the 64-Dimension Feature Vector\label{sec.FetLength}}
In our neural network (Fig.~\ref{fig.cae}b), there is only one fully connected layer, which encapsulates the extracted features of the radio galaxies in our sample. However, in typical CNNs, such as  LeNet-5 ~\citep{Lecun1995}, AlexNet~\citep{Krizhevsky2012}, and GoogLeNet~\citep{Szegedy2015}, more than one fully connected layer is employed. The reason for using only one fully connected layer in this work is that in a series of tests we find that more than one is not effective in the morphological classification tast in this work. In fact we find that by adding more than one fully connected layer, the classification precision decreases shraply instead of increasing. This phenomenon has been detected in many CNN-based works and is named degradation, i.e., the performance of the network saturates or even declines gradually with more layers than needed, (e.g.,~\citealt{He2015, He2016}). Thus, based on a series of tests, only one fully connected layer is adopted.

The length of the extracted features, i.e., the number of nodes in the fully connected layer, also affects the precision of the radio galaxy classification task and the complexity of the network, which has been discussed in many related machine learning based problems (e.g., \citealt{Krizhevsky2012,Szegedy2015,Ma2017cavity}). In Figure~\ref{fig.tsne} we illustrate the results of tree structure classification using either 64-dimension or 32-dimension vectors. We find that the degree of separation of the six subtypes (i.e., compact, typical FRI, typical FRII, BT, XRG, and RRG) obtained with 64-dimension feature vectors dominates that obtained with 32-dimension vectors, especially for the types of typical FRIs/FRIIs and BTs. This is because the shorter feature vectors squeeze the learned representations of the radio galaxy morphologies, so that some weak but useful information is lost. On the other hand, we choose not to further increase the dimensions of the features because a larger fully connected layer will exponentially increase of the scale of the neural network, which will lead to overfitting risk, affect the efficiency of the network, and thus consume more resources~\citep{Han2015,Li2016}.         

\subsection{Uncertainties in the Machine Learning Based Classification\label{sec.uncertainty}}
Two types of uncertainties, namely the aleatoric and epistemic uncertainty, are commonly discussed in machine learning based problems (e.g., \citealt{Perreault2017,Kendall2017}). The former describes the intrinsic uncertainty related to the quality of the images (e.g., occlusions, lack of visual features, and over/underexposure; see \citealt{Kendall2017} for details), which is expected to be relieved after enough high-quality images are obtained in future astronomical observations. The latter measures the uncertainty caused by the factors that the model does not take into account due to the lake of a complete training dataset. We will focus on the latter by applying the Bayesian inference as shown below.

\begin{figure}[]
\centering
\includegraphics[width=0.48\textwidth]{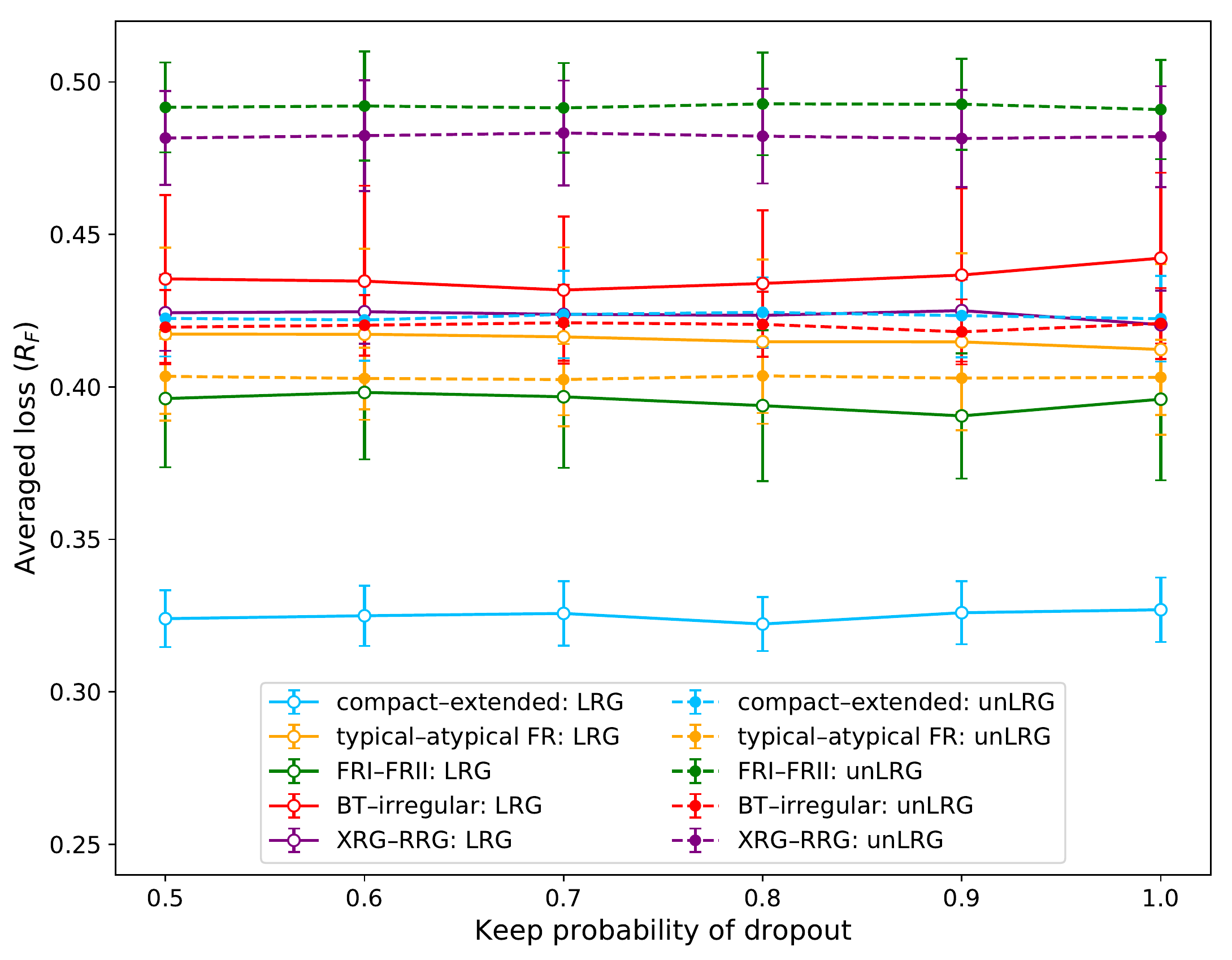}
\caption{Uncertainties of the MCRGNet on the LRG and unLRG samples as a function of the keep probability of the dropouts.\label{fig.uncern}}
\end{figure}

In the usual way, to measure the epistemic uncertainty, the prior distribution of the weight and bias parameters in the model is directly inferred, which is often referred to as the Bayesian neural network~\citep{Gal2015}. However, such direct calculations will cost a lot of time and computing resource, especially for a large-scale network. \citet{Gal2015,Kendall2017} have proved that the optimization of the neural network with dropout, which zeros out neurons randomly as a Bernoulli distribution, is approximately equivalent to a form of Bayesian inference. Thus we utilize the dropout layers already appended after each of the convolutional and fully connected layers (Section \ref{sec.AppTrnEva}) by switching them on using the approach of \citet{Gal2015}, which is called Monte Carlo (MC) dropout. Note that the dropout layers are switched off at the final stage of the labeling of the unLRG sample (see Table~\ref{tab.unLRG}). In Figure~\ref{fig.uncern} we illustrate the test loss curves together with the 68\% errors as a function of the keep probability of the dropouts for the five branches on the dichotomous tree. As expected in \S\ref{sec.DisunLRG}, for a given branch the uncertainty of the classification of the unLRG sample is in general larger than that of the LRG sample. When classifying the unLRG sample, we find again that the losses of FRI$--$FRII (green line) and XRG$--$RRG (purple line) branches are significantly higher than those of the other branches by taking the the uncertainty into consideration. This may be attributed to the unLRG sources with ambiguous morphologies (e.g., the FRIIs with relativistic beaming, which is, however, a clear feature under human inspection). These results agree with what we have obtained in \S\ref{sec.DisunLRG}. Besides, from the behavior of the error bars and the small fluctuations on the curves as the dropout keep probability decreases, we are confident that the proposed MCRGNet with dropout layers could be robust.

If we increase the threshold set to the probability, which is used to quantify how sure it is that an image belongs to a specific radio galaxy subtype (see Appendix~\ref{sec.ProbBT} for details), the precision of the predictions tends to increase, especially for the XRGs and RRGs, the numbers of which are small in both the LRG and unLRG samples (see Fig~\ref{fig.posthrs} for details). To be on the safe side we suggest an empirical probability threshold of 0.90 for selecting sources from our unLRG sample (Table~\ref{tab.unLRG}).

In addition, we have eventually propagated the uncertainty of our network (i.e., selecting the sources with the probability threshold) to Figures~\ref{fig.wlr}--\ref{fig.zwlr}, by which our conclusion remains the same as that in Section \ref{sec.cmpFR}.

\section{Summary\label{sec.summary}}
A morphological classification of 14,245 radio AGNs from the sample selected by \citet{Best2012} is performed by applying deep CNN classifiers, namely MCRGNet (the code is available at \url{https://github.com/myinxd/MCRGNet}). By combining and cross-checking the classifications by both MCRGNet and the authors' manual inspections, a catalog is also provided,  in which we retain the original items of the 14,245 radio AGNs from \citet{Best2012} and append new columns hosting the morphological and optical information obtained with SDSS.

We propose to classify the radio galaxies into six types, i.e., typical FRI/FRII, FRI/II-like BT, XRG, and RRG, where we subdivide the BT radio galaxies into FRI-like and FRII-like instead of the NAT and WAT as conventional. To realize the classification, a dichotomous tree classifier is designed, which is composed of cascaded CNN based subclassifiers. A three-step training strategy, namely PFC, is proposed to extract features of the radio galaxies for morphological classifications. Unsupervisedly pre-training of the networks with a large unlabeled sample and supervised fine-tuning with a relatively small labeled sample is proved robust to classifying radio galaxies with high precisions and sensitivities compared to results in literature (e.g.~\citealt{Aniyan2017}), as well as low epistemic uncertainty by training MCRGNet with dropout-based Bayesian inference. In addition, the relatively small-scale MCRGNet can be deployed to portable computers for the astronomers, which is a key motivation of this work.

By calculating the $r$-band absolute magnitude of the sources in the unLRG sample with the data from SDSS DR7, we find that the distributions of the unLRG sources on the $L_{\rm radio}--M_{\rm opt}$ plane, which do not show an apparent redshift evolution, indicate that there could not exist an abrupt separation between FRIs and FRIIs as reported in previous works. 

\acknowledgments
We sincerely thank the referee for providing valuable comments. We also thank Dr. H. Andernach and Dr. Y. Zhu for the revision and correction of our AGN catalog. This work was supported by the National Science Foundation of China (grant Nos. 11433002,  61371147, 11835009, 11621303, 51672176, and 11673017), and the National Key Research and Discovery Plan (grant Nos. 2018YFA0404601 and 2017YFF0210903). C.L. acknowledges the National Key Basic Research Program of China (2015CB857002). C.L. is supported by Key Laboratory for Particle Physics, Astrophysics and Cosmology, Ministry of Education, and Shanghai Key Laboratory for Particle Physics and Cosmology (SKLPPC).


%

\vspace{5mm}



\software{
Vim (\url{https://www.vim.org/}),
IPython (\url{http://ipython.org/}),
Jupyter (\url{https://jupyter.org/}),
Tensorflow (\url{https://www.tensorflow.org/}),
NumPy (\url{https://www.numpy.org/}), 
SciPy (\url{https://www.scipy.org}),
AstroPy(\url{http://www.astropy.org/};~\citealt{Astropy2018}),
Pandas (\url{https://pandas.pydata.org/}).
}

\bibliographystyle{aasjournal}
\bibliography{references}

\appendix
\section{Explanation of transposed convolution\label{sec.TransConv}}
Transposed convolution, which works by swapping the forward and backward parts of the encoder block, can be calculated by a matrix operation. As for the forwarding or the convolution of the encoder CNN, if the input $I$ and output $O$ were to be unrolled into vectors $I_v$ and $O_v$ from left to right, top to bottom, then the convolution could be represented as a sparse matrix $C$ where the nonzero elements are the elements of the weight matrix $W$. For instance, supposing the shapes of $I$, $W$, and $O$ are $3\times3$, $2\times2$, and $2\times2$ respectively, and the padding mode is same padding (i.e., half-padding) with a stride of 1, then the sparse matrix $C$ is as follows:
\begin{equation}
\label{eq.sparse}
C = 
\begin{pmatrix}
{w_{11}} & {w_{12}} & {0} & {w_{21}} & {w_{22}} & {0} & {0} & {0} & {0} \\
{0} & {w_{11}} & {w_{12}} & {0} & {w_{21}} & {w_{22}} & {0} & {0} & {0} \\
{0} & {0} & {0} & {w_{11}} & {w_{12}} & {0} & {w_{21}} & {w_{22}} & {0} \\
{0} & {0} & {0} & {0} & {w_{11}} & {w_{12}} & {0} & {w_{21}} & {w_{22}}  
\end{pmatrix}.
\end{equation}
The unrolled vectorized $I_v$ and $O_v$ of matrices $I$ and $O$ are 
\begin{align}
\label{eq.vec}
I_v  &= (I_{11}, I_{12}, I_{13}, I_{21}, I_{22}, I_{23}, I_{31}, I_{32}, I_{33} ), \\
O_v &=  (O_{11}, O_{12}, O_{21}, O_{22})^{T}.
\end{align}
Then the convolution operation is transformed into matrix multiplication,
\begin{equation}
\label{eq.conv2mul}
O_v = C \cdot I_v.
\end{equation}
Finally, by reshaping the output $O_v$ from vector to matrix, we obtain the output convolved result $O$.

Similarly, based on Equations~\ref{eq.sparse},~\ref{eq.vec}, and \ref{eq.conv2mul}, the deconvolution can be calculated by the arithmetic of transpose convolution as
\begin{equation}
\label{eq.dconv2mul}
I^{T}_v = C^{T} \cdot O^{T}_v,
\end{equation}
where the superscript $T$ represents the transpose operation; thus $C^{T}$ shares the same shape and sparsity as the transposed matrix of $C$, and so do $I^{T}_v$ and $O^{T}_v$. Finally, by reshaping $I^{T}_v$ back into $I$, we can obtain the recovered deconvolved matrix.

\section{How Does the Dichotomous Tree Classifier Work\label{sec.ProbBT}}
\begin{figure}[t]
\centering
\includegraphics[width=0.45\textwidth]{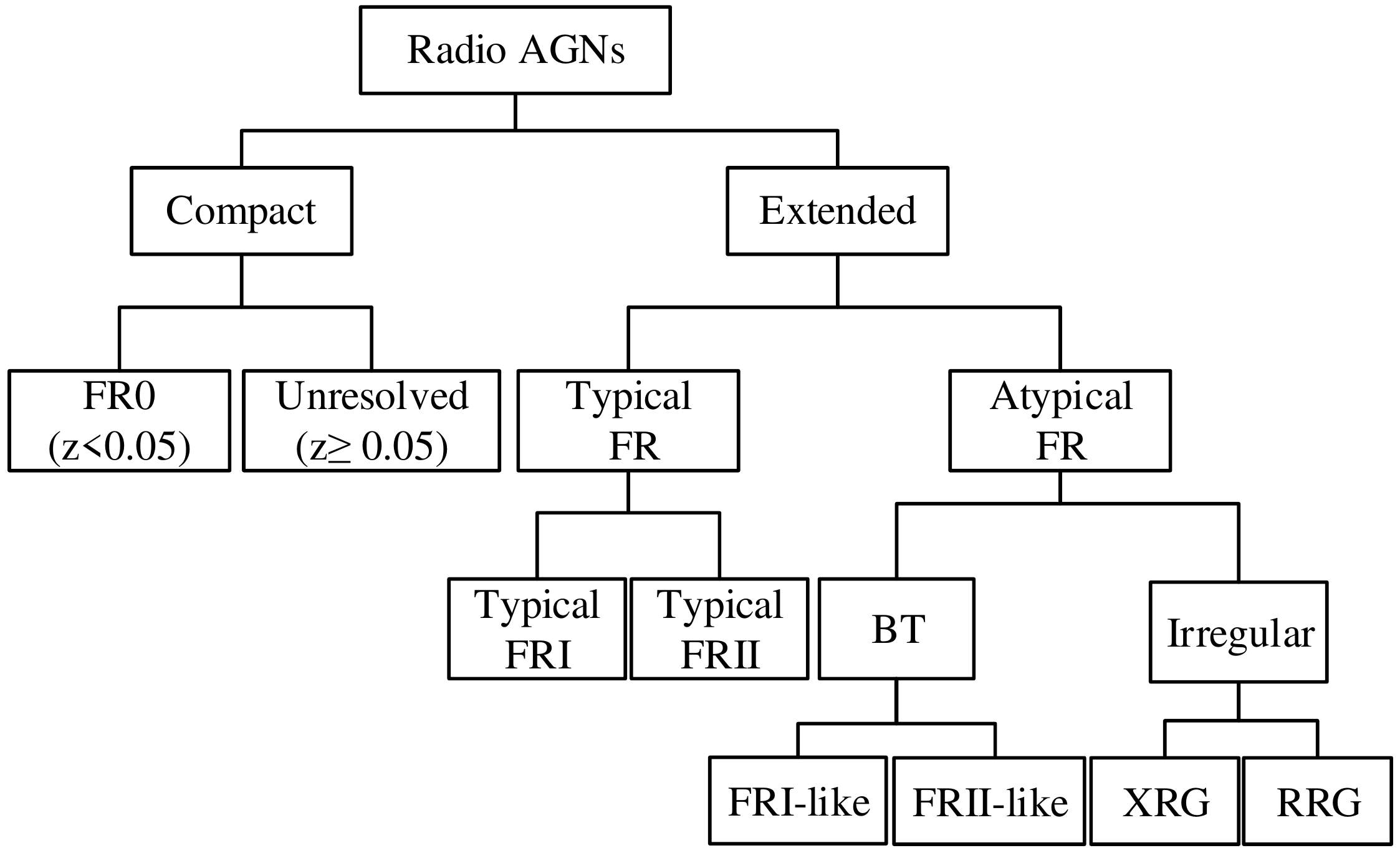}
\caption{Tree structure used to classify radio galaxies.\label{fig.tree}}
\end{figure}

\begin{figure}[]
\gridline{\fig{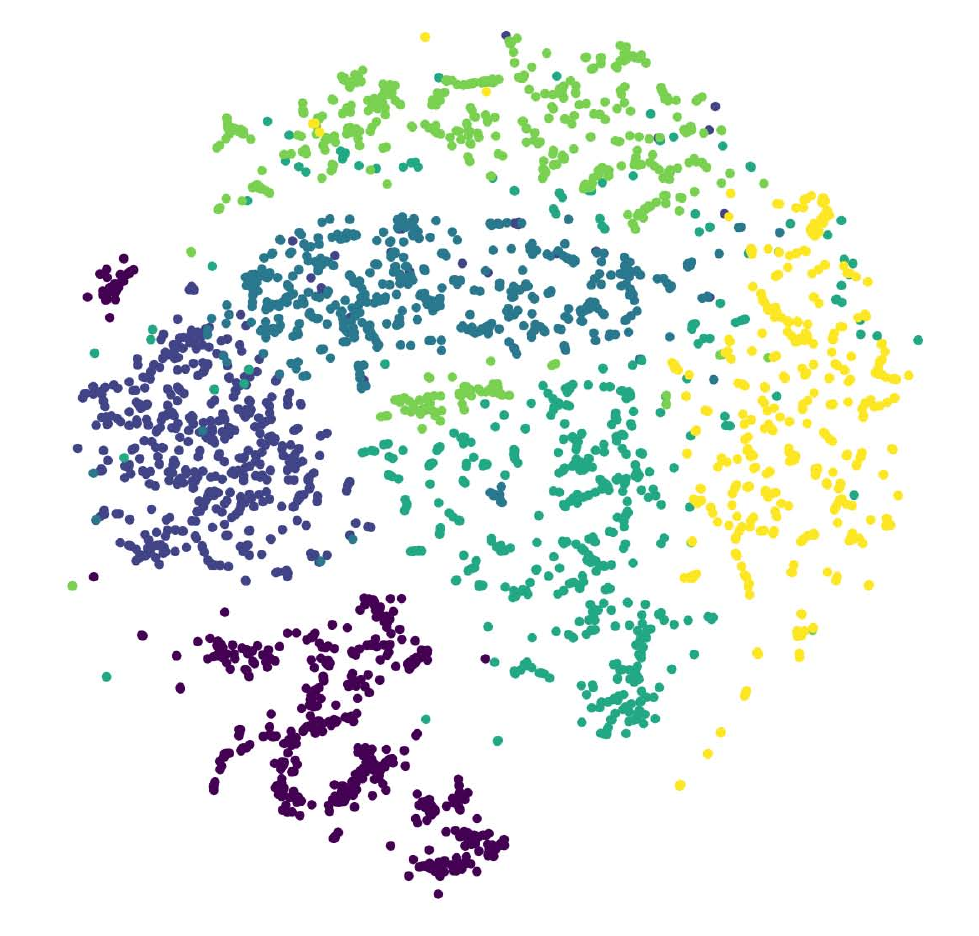}{0.36\textwidth}{(a) Obtained with 64-dimension feature vector by the dichotomous tree\label{fig.tsne64}}
          \fig{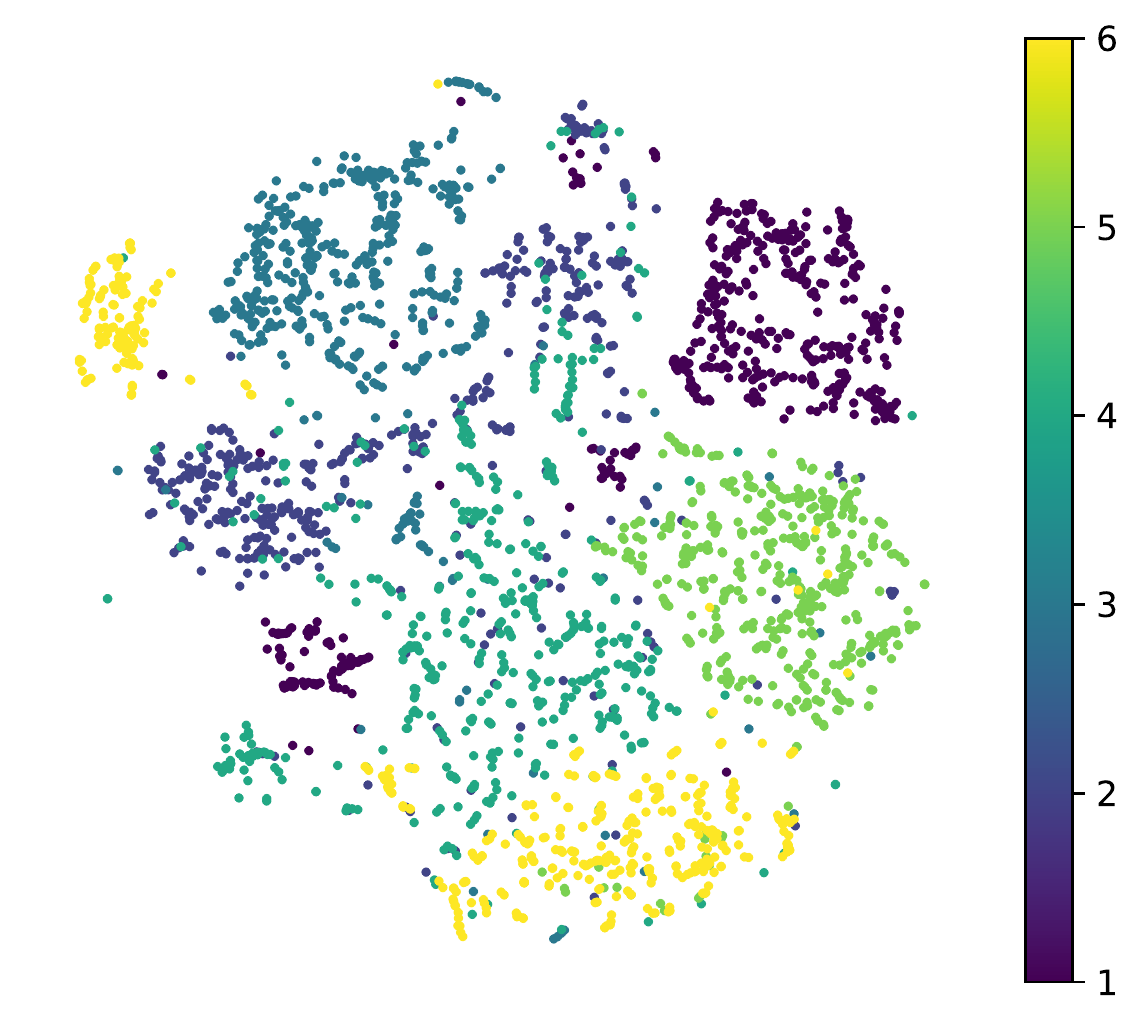}{0.35\textwidth}{(b) Obtained with 32-dimension feature vector by the dichotomous tree\label{fig.tsne32}}}
\gridline{\fig{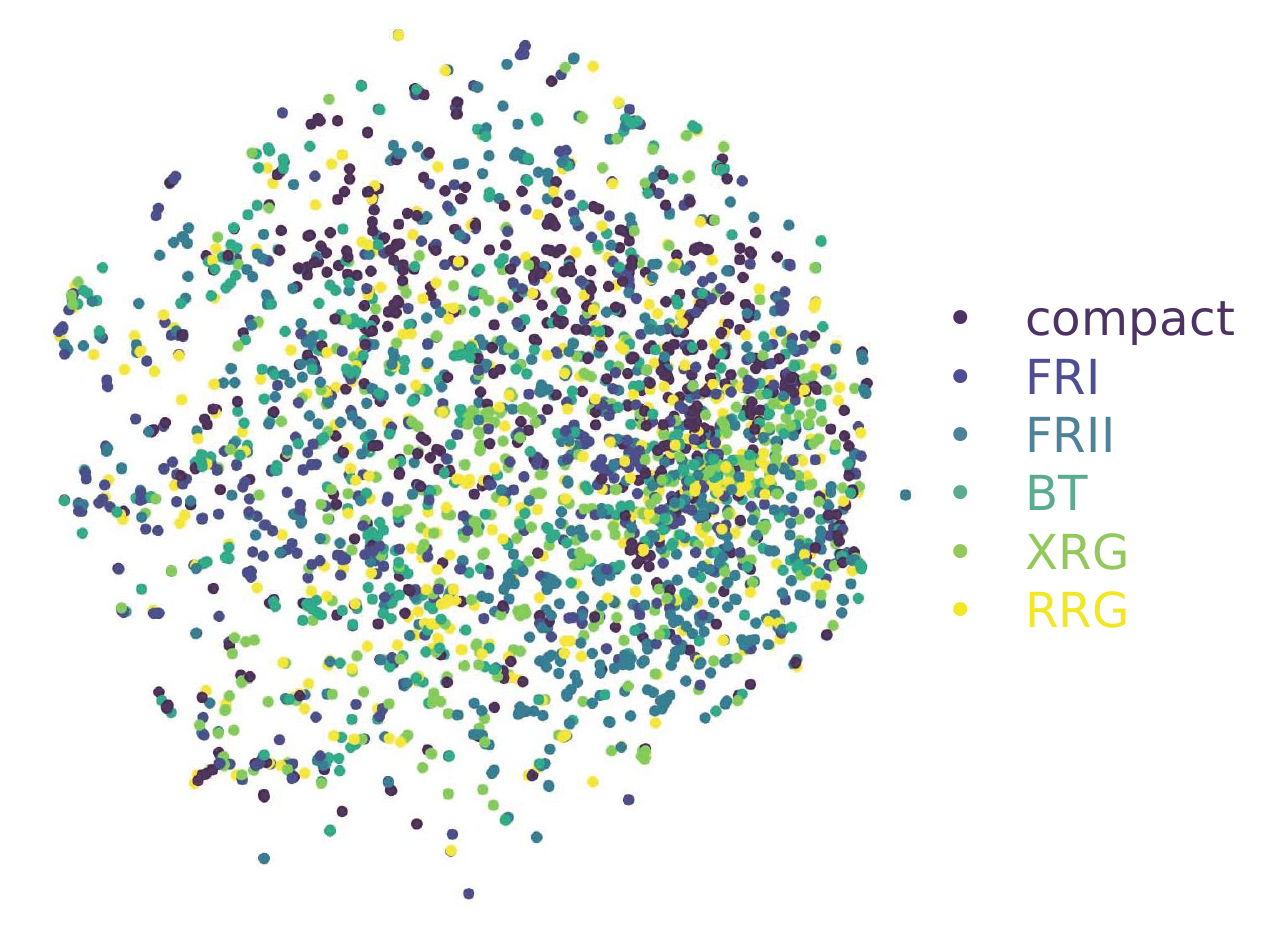}{0.5\textwidth}{(c) Obtained with 64-dimension feature vector by non-tree\label{fig.tsne_nt}}}
\caption{(a)-(b) Results of the $t$-SNE feature visualization for the classifications of the LRG samples using the dichotomous tree structure show in Fig~\ref{fig.tree}. (c) Direct six-type classification. Different types of sources are marked in different colors. 64-dimension length feature vectors are employed in (a) and (c), while 32-dimension length feature vectors in (b). \label{fig.tsne}}
\end{figure}

We propose a dichotomous tree structure to split the multiple classification for the six AGN morphological types into binary classification tasks (Fig~\ref{fig.tree}). By feeding an AGN source into the tree from the root node, a route along the tree will be obtained, and the image will arrive at one of the leaves as a certain AGN morphology. At the same time, a probability (or a reasonability as defined by \citealt{Fei2006}) would be calculated to quantify how sure it is that the tree will classify the source into a certain AGN type. Since a CNN is trained independently at each branch as described in Section \ref{sec.TrEval}, the probability of a source being classified as an AGN type (i.e., one of the leaf nodes on the tree) by the network obeys a joint distribution composed of the branches on the tree. We explain the probability calculation algorithm in this appendix, which is based on a structured probability model defined by~\citet{Goodfellow2016}.

Denote $\Br= \{\Br_1, \Br_2, ..., \Br_6\}$ as the dichotomous tree, in which $\Br_i, i=1,2,...,6$  represents the classification branches compact$--$extended, typical$--$atypical FR, typical FRI$--$FRII, BT$--$irregular, FRI-like$--$FRII-like, and XRG$--$RRG, respectively. Also, we define $\Br^{l}_i$ and $\Br^{r}_i$ as the left and right nodes of branch $i$, respectively. The joint probability is then calculated by splitting it into many factors, which are the conditional probabilities of the nodes along the decision route estimated by the corresponding CNNs, and cascadingly multiplying them together.

Following \citet{Goodfellow2016}, we define $Pa(n)$ as the parents (i.e., the former nodes of $n$ on the decision route $\mathbf{R}$) of node $n$, and
\begin{align}
P(x) &= P(\mathbf{R}) \notag \\
&=\Pi_{n\in\mathbf{R}}P_n{(n|Pa(n))},
\end{align}
where $x$ is the AGN source and $\mathbf{R}$ is the set of the nodes to be selected on the route that the source passes on the tree.

For instance, suppose we had an AGN source $x$, which was classified as an RRG by the dichotomous tree. The route on the tree would be (root) $\rightarrow$ (extended) $\rightarrow$ (atypical FR) $\rightarrow$ (irregular) $\rightarrow$ (RRG), i.e., $\mathbf{R} = \{\Br^r_1, \Br^r_2,\Br^r_4, \Br^r_6\}$). Then the joint probability of x is calculated by,
\begin{align}
P(x) &= P(\mathbf{R}=\{\Br^r_1, \Br^r_2,\Br^r_4, \Br^r_6\}) \notag \\
        &= \Pi_{n\in\mathbf{R}}P_n{(n|Pa(n))} \notag \\
        &= P(\Br^r_1)P(\Br^r_2|\Br^r_1)P(\Br^r_4|\Br^r_2,\Br^r_1)P(\Br^r_6|\Br^r_4,\Br^r_2,\Br^r_1).
\end{align}

We provide the joint probability as the reasonability for each of the sources in both the LRG and unLRG samples obtained by classifying them with the trained tree in Tables ~\ref{tab.LRG} and \ref{tab.unLRG}. To utilize the unLRG catalog in further researches, we suggest an empirical probability threshold of 0.90 as a safe reference for the users to select their AGN samples with (see Section \ref{sec.DisunLRG} for a detailed discussion).

\newpage


\begin{deluxetable}{lp{3cm}p{2cm}}[t]
\tabletypesize{\footnotesize}
\tablecaption{Catalogs Used to Construct the LRG Sample.\label{tab.tab1}}
\setlength{\tabcolsep}{9pt}
\tablehead{
\colhead{Catalog} & 
\colhead{$^{\rm a}$Source Number \& Type} & 
\colhead{Reference}}
\startdata
FR0CAT  & 108 (compact: 104) & \citet{Baldi2017} \\
FRICAT	 & 233 (FRI: 192; compact: 1; BT: 5) & \citet{Capetti2017a} \\
FRIICAT  & 123 (FRII: 80; BT: 8; XRG: 3)	 & \citet{Capetti2017b} \\
\citet{Cheung2007} & 100 (XRG: 81) & \citet{Cheung2007} \\
\citet{Proctor2011} & 475 (BT: 285; RRG: 32) & \citet{Proctor2011} \\
CoNFIG 1--4  & 856 (FRI: 19; FRII: 351; BT: 9; compact: 272) & \citet{Gendre2010} \\
\enddata
\tablecomments{Sources marked as unconfirmed or uncertain in the CoNFIG catalog are not included in this work. Sources with low PAPRs or spanning a region larger than the $150\times150$ pixel crop box are also excluded.}
\tablenotetext{{\rm a}}{Source number in the original catalogs and source number included in our LRG sample with corresponding types (brackets).}
\end{deluxetable}

\begin{deluxetable}{lccr}[t]
\tablecaption{Number of Sources Included in the LRG Sample.\label{tab.tab2}}
\setlength{\tabcolsep}{12pt}
\tablehead{
\colhead{AGN { Type}} & 
\colhead{$^{\rm a}$Label} & 
\colhead{Source Number} & 
\colhead{$^{\rm b}$Catalog(s)}
}
\startdata
compact & 1 & 377 & 1,2,3 \\
FRI & 2 & 211 & 1,3 \\
FRII & 3 & 431 & 1, 3, 4 \\
{ BT} & 4 & 307 & 1, 3, 4, 5 \\
XRG & 5 & 84 & 6 	\\
RRG & 6 & 32 & 5 	\\ \hline
Total & $\cdots$ & 1442 & $\cdots$ \\
\enddata
\tablenotetext{{\rm a}}{Labels in this work of corresponding AGN types.}
\tablenotetext{{\rm b}}{Catalogs used to compose the LRG sample: (1) CoNFIG, (2) FR0CAT, (3) FRICAT, (4) FRIICAT, (5) \citet{Proctor2011}, and (6) \citet{Cheung2007}}
\end{deluxetable}

\begin{deluxetable*}{lcccccccccc}[t]
\tablecaption{1442 LRGs in the LRG Sample.\label{tab.LRG}}
\tablehead{
\colhead{Name} & 
\colhead{{R.A.}}  &  
\colhead{{Decl.}} & 
\colhead{ $z$} & 
\colhead{$S_{\rm 1.4~GHz}^{\rm{NVSS}}$} &
\colhead{$^{\rm a}$Catalog} & 
\colhead{BH12} & 
\colhead{$^{\rm b}$Type} & 
\colhead{$^{\rm b}$Type} & 
\colhead{BT} & 
\colhead{ Probability} \\ 
\colhead{} &
\colhead{(h)} &
\colhead{(deg)} &
\colhead{} & 
\colhead{(mJy)} & 
\colhead{} & 
\colhead{} & 
\colhead{} &
\colhead{} & 
\colhead{} \\
\colhead{(J2000)} &
\multicolumn{2}{c}{(J2000)} &  
\colhead{} &
\colhead{} &
\colhead{(Literature)} &
\colhead{} &
\colhead{(Literature)} &
\colhead{(CNN)} &
\colhead{(Subtype)}
}
\startdata
J$000140.17-003350.6$&   0.02783&  $-0.56406$& 0.2469&   73.00& 6& 0&  5&  5& N/A& 0.97044\\
J$000330.73+002756.1$&   0.05854&   $0.46558$&    N/A&     N/A& 5& 1&  4&  4&   1& 0.99282\\
J$001247.57+004715.8$&   0.21321&   $0.78772$& 0.1480&   58.60& 4& 1&  3&  3& N/A& 1.00000\\
J$002107.62-005531.4$&   0.35212&  $-0.92539$& 0.1080&  111.90& 4& 1&  3&  3& N/A& 0.99320\\
J$002900.98-011341.7$&   0.48361&  $-1.22825$& 0.0830&  282.80& 3& 1&  1&  1& N/A& 1.00000\\
J$003302.41-014956.6$&   0.55067&  $-1.83239$& 0.1301&   87.00& 6& 0&  5&  5& N/A& 0.93146\\
J$003636.21+004853.4$&   0.61006&   $0.81483$&    N/A&  280.00& 6& 0&  5&  5& N/A& 0.95816\\
J$003930.52-103218.6$&   0.65848& $-10.53850$& 0.1290&   23.70& 3& 1&  2&  2& N/A& 1.00000\\
J$004012.89+012542.1$&   0.67025&   $1.42836$&    N/A&     N/A& 5& 0&  4&  3& N/A& 0.95473\\
J$004148.22-091703.1$&   0.69673&  $-9.28419$& 0.0530&   53.00& 3& 1&  2&  2& N/A& 0.96172\\
\enddata
\tablecomments{The Columns show (1) the source names (J2000), (2)--(4) the sky coordinates and redshifts (samples without redshift records are marked with N/A), (5) the 1.4 GHz flux densities obtained with the NVSS data, (6) the previous catalogs in literature used in this work, (7) flags of the sources (1 means included and 0 means not included in the BH12 sample (\citealt{Best2012}), (8)--(9) the source types labeled in literature and by our CNN, (10) the labels for the BT sample (1 = FRI-like source, 2 = FRII-like source, and N/A = non-BT source), (11) the probabilities of the classifications. Only the first ten sources are listed here, and the full table is available electronically.}
\tablenotetext{\rm a}{Catalogs in literature for radio galaxies: (1) CoNFIG, (2) FR0CAT, (3) FRICAT, (4) FRIICAT, (5) \citet{Proctor2011}, (6) \citet{Cheung2007}.}
\tablenotetext{\rm b}{Labels of { radio galaxy} types in this work: (1) compact sources (FR0s are marked as $1_{\rm F}$), (2) typical FRIs, (3) typical FRIIs, (4) BTs, (5) XRGs, and (6) RRGs.}
\end{deluxetable*}

\begin{deluxetable*}{lrrrrrrrrrr}[t]
\tablecaption{14,245 Unlabeled Radio AGNs (unLRG) in the unLRG Sample.\label{tab.unLRG}}
\tablehead{
\colhead{Name}  & 
\colhead{ R.A.}   &  
\colhead{ Decl.} & 
\colhead{z} & 
\colhead{$S_{\rm 1.4~GHz}^{\rm{NVSS}}$}  & 
\colhead{$M^{\rm{SDSS}}_r$} & 
\colhead{$^{\rm a}$Type} & 
\colhead{$^{\rm b}$BT} & 
\colhead{$^{\rm a}$Type} & 
\colhead{$^{\rm b}$BT} & 
\colhead{ Probability} \\
\colhead{} &
\colhead{(h)} &
\colhead{(deg)} & 
\colhead{} & 
\colhead{(mJy)} & 
\colhead{(mag)} &
\colhead{} & 
\colhead{} &
\colhead{} & 
\colhead{} & 
\colhead{} \\
\colhead{(J2000)} &
\multicolumn{2}{c}{(J2000)} &  
\colhead{} &
\colhead{} &
\colhead{} &
\colhead{(CNN)} &
\colhead{(CNN)} &
\colhead{(MAN)} &
\colhead{(MAN)} 
}
\startdata
J000001.57$-$092940.3&   0.00044&  $-9.49453$& 0.1905&    9.10& $-22.45$&  1& N/A&  1& N/A& 0.99992\\
J000025.55$-$095752.8&   0.00710&  $-9.96467$& 0.3540&   14.10& $-22.21$&  2& N/A&  2& N/A& 1.00000\\
J000027.89$-$010235.4&   0.00775&  $-1.04317$& 0.4388&   12.60& $-23.26$&  1& N/A&  1& N/A& 1.00000\\
J000049.32$-$005042.9&   0.01370&  $-0.84525$& 0.4110&   22.70& $-22.97$&  2& N/A&  3& N/A& 0.96863\\
J000052.92$+$003044.6&   0.01470&   $0.51239$& 0.3398&   15.70& $-22.21$&  3& N/A&  3& N/A& 0.91637\\
J000052.99$-$011020.4&   0.01472&  $-1.17233$& 0.1948&   11.20& $-22.26$&  2& N/A&  3& N/A& 0.95235\\
J000121.46$-$001140.3&   0.02263&  $-0.19453$& 0.4615&  110.90& $-22.49$&  4&   1&  4&   1& 0.99931\\
J000121.53$+$010149.0&   0.02265&   $1.03028$& 0.5520&   18.10& $-22.47$&  2& N/A&  3& N/A& 0.92581\\
J000134.89$-$085154.8&   0.02636&  $-8.86522$& 0.1765&   50.00& $-22.52$&  3& N/A&  3& N/A& 0.99955\\
J000134.97$-$085727.7&   0.02638&  $-8.95769$& 0.1737&    5.00& $-22.49$&  1& N/A&  1& N/A& 1.00000\\
\enddata
\tablecomments{{ The columns show} (1) the source names (J2000), (2)--(4) the sky coordinates and redshifts, (5) the 1.4 GHz flux densities obtained with the NVSS data, (6) the $r$-band absolute magnitudes obtained with the SDSS data, (7)--(10) the source types labeled by our CNN network and our manual inspection, and (11) the probabilities of the classifications. Only the first 10 sources are listed here, and the full table is available electronically.}
\tablenotetext{\rm a}{Labels of radio galaxy types determined in this work with CNN (column 7) and manually (column 9): (1) compact sources (FR0s are marked as $1_{\rm F}$), (2) typical FRIs, (3) typical FRIIs, (4) BTs, (5) XRGs, and (6) RRGs.}
\tablenotetext{\rm b}{Labels of BT subtypes determined in this work with CNN (column 8) and manually (column 10): (1) FRI-like sources, (2) FRII-like sources, and (N/A) non-BT sources.}
\end{deluxetable*}

\begin{deluxetable*}{lrrrrr}
\tablecaption{The design of the CAE network and the CNN classifier.\label{tab.cae}}
\setlength{\tabcolsep}{7pt}
\tablehead{
\colhead{Layer} & 
\colhead{$^{\rm a}$Kernel Size} & 
\colhead{$^{\rm b}$Channels} & 
\colhead{$^{\rm c}$Padding} & 
\colhead{$^{\rm d}$Stride} & 
\colhead{$^{\rm e}$Subnet}
}
\startdata
Conv-En1  & $3\times3$ & 8   & same & 2 	 & Encoder and CNN \\
Conv-En2  & $3\times3$ & 8   & same & 2 	 & Encoder and CNN \\
Conv-En3  & $3\times3$ & 16  & same & 2 	 & Encoder and CNN \\
Conv-En4  & $3\times3$ & 16  & same & 2 	 & Encoder and CNN \\
Conv-En5  & $3\times3$ & 32  & same & 2 	 & Encoder and CNN \\
Encode    & 64         & --- & ---  & ---    & Encoder \& CNN \\
Softmax   & 2          & --- & ---  & --- & CNN \\
Conv-De1 & $3\times3$ & 32  & same & 2 	 & Decoder \\
Conv-De2 & $3\times3$ & 16  & same & 2	 & Decoder \\
Conv-De3 & $3\times3$ & 16  & same & 2 	 & Decoder \\
Conv-De4 & $3\times3$ & 8   & same & 2	 & Decoder \\
Conv-De5 & $3\times3$ & 8   & same & 2 	 & Decoder \\
\enddata
\tablenotetext{{\rm a}}{Kernel sizes are set set to 3 $\times$ 3 pixels according to AlexNet~\citep{Krizhevsky2012} and VGG~\citep{Simonyan2014}.}
\tablenotetext{{\rm b}}{Number of the channels with respect to the layer.}
\tablenotetext{{\rm c}}{Padding modes for the convolution or deconvolution arithmetic.}
\tablenotetext{{\rm d}}{Stride modes for the convolution or deconvolution arithmetic.}
\tablenotetext{{\rm e}}{Subnet of the CAE to which the layer belongs.}
\end{deluxetable*}

\begin{deluxetable}{lrrrrrr}
\tablecaption{Augmentations of the LRG sample.\label{tab.aug}}
\setlength{\tabcolsep}{7pt}
\tablehead{
\colhead{AGN Type} & 
\colhead{Label} & 
\colhead{$^{\rm a}N_{\rm Trn+Val}$} & 
\colhead{$^{\rm b}N_{\rm Tst}$} & 
\colhead{$^{\rm c}R_{\rm Aug}$} & 
\colhead{$^{\rm d}N_{\rm Aug}$} &
\colhead{$^{\rm e}$Branch}
}
\startdata
compact 	& 1 	& 302 	& 75 	 & 64 	& 19,328   & 1\\
FRI 		& 2 	& 169 	& 42 	& 29 	&  4901 & 1,2,3\\
FRII 		& 3 	& 345 	& 86 	& 14 	& 4830 & 1,2,3	 \\
BT 		& 4 	& 245 	& 62 	& 20 	& 4900 & 1,2,4,5\\
XRG 	& 5 	& 67 	& 17 	& 37 	& 2479	 & 1,2,4,6\\
RRG 	& 6 	& 26 	& 6 		& 94 	& 2444 & 1,2,4,6 \\ \hline
Total 	& --- 	& 1154 	& 288 	& ---  	& 38,882 & --- \\
\enddata
\tablenotetext{{\rm a}}{Numbers of the sources in the training (Trn) + validation (Val) subsets.}
\tablenotetext{{\rm b}}{Numbers of the sources in the test (Tst) subsets.}
\tablenotetext{{\rm c}}{Augmentation rates for the corresponding RG types.}
\tablenotetext{{\rm d}}{Numbers of the augmented images for corresponding RG types.}
\tablenotetext{{\rm e}}{Branches on the dichotomous tree: (1) compact--extended, (2) typical--atypical FR, (3) typical FRI--FRII, (4) BT--irregular, (5) FRI-like--FRII-like, and (6) XRG--RRG.}
\end{deluxetable}

\begin{deluxetable}{lrrrrrrrrr}
\tablecaption{Evaluation of the CNN network based on the classification of the LRG sample.\label{tab.eval}}
\setlength{\tabcolsep}{5pt}
\footnotesize
\tablehead{
\colhead{AGN Type} & 
\multicolumn{3}{c}{Pre-trained} &
\multicolumn{3}{c}{Non-pre-trained} &
\multicolumn{3}{c}{Gain}  \\
\colhead{} &
\colhead{$R_{\rm sen}$} & 
\colhead{$R_{\rm acc}$} & 
\colhead{$R_{\rm pre}$} &
\colhead{$R_{\rm sen}$} & 
\colhead{$R_{\rm acc}$} & 
\colhead{$R_{\rm pre}$} &
\colhead{$R_{\rm sen}$} & 
\colhead{$R_{\rm acc}$} & 
\colhead{$R_{\rm pre}$} \\
\colhead{} & 
\multicolumn{3}{c}{(\%)} & 
\multicolumn{3}{c}{(\%)} & 
\multicolumn{3}{c}{(\%)} 
}
\startdata
compact & 
$98.41\pm0.25$ & $98.96\pm0.16$ & $97.88\pm0.63$ &  
$98.14\pm0.27$ & $98.82\pm0.17$ & $97.12\pm0.67$ &
$0.27\pm0.17$ & $0.14\pm0.13$ & $0.76\pm0.58$ \\
FRI & 
$88.72\pm1.69$ & $96.94\pm0.48$ & $95.39\pm2.73$ &
$83.46\pm2.13$ & $95.42\pm0.56$ & $93.94\pm2.74$ &
$5.26\pm1.32$  & $1.52\pm0.57$ & $1.45\pm2.80$\\
FRII &
$91.07\pm2.12$ & $95.66\pm0.39$ & $92.24\pm1.01$ &
$80.77\pm1.97$ & $88.49\pm0.46$ & $88.41\pm1.48$ &
$10.90\pm1.52$ & $7.27\pm0.33$ & $3.83\pm1.27$\\
BT &
$94.27\pm1.89$ & $95.25\pm0.46$ & $86.56\pm0.92$ &
$85.57\pm2.61$ & $89.88\pm0.52$ & $76.57\pm1.88$ &
$8.70\pm1.32$ & $5.37\pm0.43$ & $9.99\pm1.50$ \\
XRG &
$75.05\pm3.54$ & $96.72\pm0.34$ & $72.34\pm2.70$ &
$64.29\pm4.78$ & $91.19\pm0.53$ & $55.76\pm4.27$ &
$10.78\pm2.44$ & $5.61\pm0.58$ & $16.58\pm4.32$ \\
RRG &
$74.62\pm3.64$ & $97.04\pm0.30$ & $65.89\pm3.65$ &
$70.42\pm6.47$ & $95.15\pm0.28$ & $60.21\pm5.88$ & 
$4.20\pm4.52$ & $1.89\pm0.45$ & $5.68\pm7.66$\\ \hline
Total &
--- & --- & $92.44 \pm 1.94 $&
--- & --- & $84.67 \pm 2.82 $&
--- & --- & $7.77 \pm 3.02 $\\
\enddata
\tablecomments{$R_{\rm sen}$, $R_{\rm acc}$, and $R_{\rm pre}$ are the evaluation indices, namely the sensitivity, accuracy, and precision, of the corresponding type.}
\end{deluxetable}

\begin{deluxetable}{lrrrr}
\tablecaption{Evaluation of the CNN network based on the comparison between the results obtained with the CNN and manual classification of the unLRG sample.\label{tab.unLRG_vis}}
\setlength{\tabcolsep}{7pt}
\tablehead{
\colhead{AGN type} & 
\dcolhead{^{\rm a}N_{\rm CNN}} & 
\dcolhead{^{\rm b}N_{\rm crt}} & 
\dcolhead{^{\rm c}N_{\rm man}} & 
\dcolhead{^{\rm d}R_{\rm pre}} \\
\colhead{} &  
\colhead{} & 
\colhead{} & 
\colhead{} & 
\colhead{(\%)} 
}
\startdata
compact 	& 5572  	& 5422 	& 6108    & 88.86 \\
FRI 		& 4781 	& 4032 	& 5048 	& 79.87  \\
FRII 	         & 2299 	& 1233    & 2083 	& 59.19  \\
BT 		& 1218 	& 535 	& 886 	& 60.38 \\
XRG 	& 290 	& 36 	& 78 	& 46.15 \\
RRG 	& 85 	& 17 	& 48 	& 35.52 \\ \hline
Total 	& {14,245} 	& 11,275  & 14,245 & 79.15\\
\enddata
\tablenotetext{{\rm a}}{Numbers of the sources classified by the CNN for the corresponding types.}
\tablenotetext{{\rm b}}{Numbers of the sources correctly classified by the CNN.}
\tablenotetext{{\rm c}}{Numbers of the sources manually classified for the corresponding types.}
\tablenotetext{{\rm d}}{Precisions of the CNN classifications calculated by comparing with human inspections.}
\end{deluxetable}

\end{document}